\documentclass[aip,graphicx,amsmath,amssymb,cha]{revtex4-1}

\usepackage{graphicx}

\draft 

\begin{document}


\title{On the origin of chaotic attractors with two zero Lyapunov exponents in a system of five biharmonically coupled phase oscillators}



\author{Evgeny A. Grines}
 \email{evgenij.grines@gmail.com}
\affiliation{Lobachevsky   State   University   of   Nizhni   Novgorod,   23
Gagarin av., Nizhny Novgorod 603950, Russia}
\author{Alexey Kazakov}
 \email{kazakovdz@yandex.ru}
\affiliation{National Research University Higher School of Economics,
25/12 Bolshaya Pecherskaya Ulitsa, 603155 Nizhny Novgorod, Russia}
\author{Igor R. Sataev}
 \email{sataevir@gmail.com}
\affiliation{Kotelnikov’s Institute of Radio-Engineering and Electronics of RAS, Saratov Branch, Zelenaya 38, Saratov,
410019, Russia}


\date{\today}

\begin{abstract}

We study chaotic dynamics in a system of four differential equations describing the dynamics of five identical globally coupled phase oscillators with biharmonic coupling. We show that this system exhibits strange spiral attractors (Shilnikov attractors) with two zero (indistinguishable from zero in numerics) Lyapunov exponents in a wide region of the parameter space. We explain this phenomenon by means of bifurcation analysis of the three-dimensional Poincar\'e map for the system under consideration. We show that the chaotic dynamics develop here near a codimension three bifurcation, when a periodic orbit (fixed point in the Poincar\'e map) has the triplet $(1, 1, 1)$ of multipliers. As it is known, the asymptotic flow normal form for this bifurcation coincides with the three-dimensional Arneodo-Coullet-Spiegel-Tresser (ACST) system in which spiral attractors exist. Based on this, we conclude that the additional near-zero Lyapunov exponent for orbits in the observed attractors appear due to the fact that the corresponding three-dimensional Poincar\'e map is close to the time-shift map of the three-dimensional ACST-system.

\end{abstract}


\maketitle 

\begin{quotation}
It is well-known that chaotic attractors of three-dimensional systems of differential equations always have in numerical experiments one positive, one zero (indistinguishable from zero) and one negative Lyapunov exponents. In four-dimensional systems of differential equations two types of chaotic attractors are well-known: hyperchaotic attractors possessing two positive, one zero, and one negative Lyapunov exponents and strongly dissipative attractors which have one positive, one zero, and two negative Lyapunov exponents. This paper deals with the quite recently discovered phenomenon when chaotic attractors of a four-dimensional system of differential equations demonstrate two zero Lyapunov exponents in a large noticeable region of parameter space. As a case study we consider the system describing dynamics of five identical globally coupled phase oscillators with biharmonic coupling. Analysing bifurcations leading to the appearance of chaotic attractors in this system, we conclude that these attractors get the additional zero Lyapunov exponents due to the fact that their 3D Poincar\'e map can be well-approximated by a time-shift map of a three-dimensional system of differential equations. The attractors under consideration include the so-called discrete Shilnikov attractors that have, for the 3D Poincar\'e map, a saddle-focus fixed point and contain entirely its two-dimensional unstable invariant manifold. It would seem that Shilnikov attractors should also be hyperchaotic \cite{GonGonShil2012, GSKK19}. However, as it is also shown in our numerics, this is not true for the system under consideration. Finally, we stress that our explanation is also suitable for other systems demonstrating chaotic attractors with the additional zero Lyapunov exponent.
\end{quotation}

\section*{Introduction}

Systems of interacting elements are of great interest to specialists in nonlinear dynamics and dynamical systems. The collective behaviour of such systems can be quite complex \cite{PikRos2015,StankPerMcClintStefa2017} even in the case when dynamics of a single element are very simple. One of approaches to the study of such systems is connected with their approximation by systems of coupled phase oscillators, whose individual dynamics are governed by equation $\dot \theta_n = \omega_n$.

Systems of identical phase oscillators ($\omega_n =\omega$ for all $n$) are of special interest here. In systems of identical elements, the type of coupling between oscillators becomes one of the main sources of complex behaviour. For example, choosing the coupling function to have only the first Fourier harmonic results in the Kuramoto~\cite{Kur1975} or Kuramoto-Sakaguchi system~\cite{KurSak1986}. Although these systems are widely used to study and explain synchronization, not all phenomena of collective behavior can be observed in them; for example, such systems cannot have several clusters of synchronized elements \cite{EngelMir2014,PikRos2015}. However, adding the second harmonic to the coupling function makes dynamics much richer, allowing formation of clusters of synchronizations \cite{Okuda1993}, heteroclinic cycles \cite{HansMatoMeunier1993,kori2001slow} and chaotic attractors \cite{AshTownOrosz2007}. Note that in the case of non-identical oscillators, chaotic dynamics can emerge even in the simplest Kuramoto models due to the detuning of the frequencies $\omega_n$ of oscillators \cite{PopMaistrTass2005}. There are also other ways to get chaotic attractors in systems of identical globally coupled phase oscillators. For example, this could be done by using the coupling function possessing higher order harmonics \cite{AshBickTimm2011} or when this function depends on more than two phase variables \cite{AshBickRodr2016}. Thus, since we have examples of chaotic attractors in systems of identical globally coupled phase oscillators \cite{Okuda1993, PopMaistrTass2005, AshTownOrosz2007, AshBickTimm2011, AshBickRodr2016}, it is very natural to ask: how do chaotic attractors in these systems appear and what are they? 

The present work studies the origin and nature of chaotic dynamics in a four-dimensional ODE system \eqref{eq_mainEq} describing interaction of five identical globally coupled phase oscillators with biharmonic coupling (this system has the effective dimension four for the phase differences). Strange attractors in this system were found in Ref.~\cite{AshTownOrosz2007}. We show that this system belongs to a very specific class of dynamical systems where chaotic attractors have two zero (indistinguishable from zero in numerical experiments) Lyapunov exponents. Moreover, this property persists in a sufficiently large region of the parameter space. Systems with such a behavior are known, see e.g. Refs.~\cite{broer2002bifurcations, broer2005quasi, GOST05, broer2010chaos, stankevich2020three, SKKS20, KSK21}, however reasons for the appearance of such chaotic attractors in many cases are not clear. As far as we know, the corresponding explanation is given only in the following two cases: (i) when a system is subjected to an external quasiperiodic or skew forcing, see e.g. Ref.~\cite{broer2010chaos}; (ii) when a system has a lower dimension due to the presence of additional invariants, for example, first integral \cite{GGK13, BorKazSat2016} (iii) when chaotic attractors exist in some neighborhood of a codimension-three bifurcation in the parameter space \cite{GOST05, SKKS20, KSK21}. 

We perform a detailed bifurcation analysis and conclude that the additional near-zero Lyapunov exponent for chaotic attractors of system \eqref{eq_mainEq} appears due to the fact that the three-dimensional Poincar\'e map for the system can be well approximated by a time shift map of a certain three-dimensional system of differential equations. Namely, we find that chaotic dynamics in the system develop near the codimension three bifurcation, when a periodic orbit has the triplet $(1,1,1)$ of multipliers. An asymptotic normal form for this bifurcation is the Arneod\'o-Coullet-Spiegel-Tresser (ACST) system \cite{ArnColTreSp1985,ArnColTreSp1985b} (see formula \eqref{eq_ACT} below). As was shown in Refs.~\cite{ArnCoulTres1982,ArnColTreSp1985,ArnColTreSp1985b}, bifurcations of the ACST-system can lead to the appearance of Shilnikov spiral attractors containing both a saddle-focus equilibrium, its two-dimensional invariant manifold, and a homoclinic orbit (called also a homoclinic loop of the saddle-focus). Therefore, in the 3D Poincar\'e map of the initial 4D system \eqref{eq_mainEq}, one can observe discrete analogues of such attractors -- the so-called discrete Shilnikov attractors \cite{GonGonShil2012, GonGonKazTur2014, GonGon2016} containing a saddle-focus fixed point and its entire two-dimensional unstable manifold. Since the orbits of the 3D Poincar\'e map ``follow'' orbits of continuous asymptotic normal form, one can expect the appearance of very close to zero Lyapunov exponent for the Poincar\'e map and, as a result, a pair of indistinguishable from zero Lyapunov exponents in the initial four-dimensional system \eqref{eq_mainEq}.

The paper is organized as follows. In Section~\ref{sec1}, we introduce the system under study, discuss main dynamical regimes, and show that chaotic attractors observed in this system have a pair of indistinguishable from zero Lyapunov exponents. In Section~\ref{sec2}, we present two-parameter bifurcation analysis of the system and give explanation of this phenomenon. In Section~\ref{sec3}, we show that one of the most interesting attractors in the system are discrete Shilnikov attractors. We describe bifurcation scenarios leading to these attractors in a one-parameter family and give numerical evidence that these attractors contain a saddle-focus fixed point together with its unstable two-dimensional manifold and homoclinic orbits. It is interesting to note that the stable and unstable manifolds of the fixed point belonging to the observed Shilnikov attractors almost do not split (see Fig.~\ref{fig7}b) which is not typical for 3D maps. We discuss this phenomenon in Section~\ref{sec4}. This paper is a development of ideas and observations from our earlier draft \cite{GrinKazSat2017}.

\section{Main equations and dynamical regimes}\label{sec1}

In paper\cite{AshTownOrosz2007} the following system of $N$ phase oscillators was considered: 
$$ \dot{\theta}_n = \omega_n + \frac{1}{N}\sum\limits_{m=1}^{N} g(\theta_n - \theta_m) + \epsilon I_n(t) + \eta w_n(t),\, n=1,\dots, N,$$
where $\theta_n$ is the phase of $n$-th oscillator, $\omega_n$ is its natural frequency, $I_n (t)$ is an impulsive input with unit magnitude, $w_n(t)$ is uncorrelated white noise such that the associated random walk has unit growth of variance per unit time, and $g(\theta)$ is the coupling function.
If we suppose that all oscillators are identical (i.e., $\omega_n = \omega$ for all $n=1, \dots, N$) and there are no both the external inputs and noises, then the system takes the form 
$$ \dot{\theta}_n = \omega + \frac{1}{N}\sum\limits_{m=1}^{N} g(\theta_n - \theta_m), \,n=1,\dots, N.$$
For any number $N$ of oscillators the dimension of this system can be reduced by one \cite{AshSwift1992} if to rewrite it for the phase differences $\gamma_n = \theta_n - \theta_{n^\ast}$, where $n^\ast$ is any number from $\{ 1, \dots, N \}$. The same as in Ref.~\cite{AshTownOrosz2007} we consider the case of five oscillators, $N=5$, with the biharmonic coupling function of the form $g(\theta) = -\sin{(\theta + \alpha)} + r \sin{(2\theta + \beta)}$, and take $n^\ast = 5$. The corresponding system for phase differences $\gamma_n = \theta_n - \theta_5$ then takes the form
\begin{equation}
\label{eq_mainEq}
\dot{\gamma}_n = \Gamma_n (\gamma_1, \gamma_2, \gamma_3, \gamma_4) =  \frac{1}{5}\sum\limits_{m=1}^4 \left \lbrack g(\gamma_n - \gamma_m) -g(-\gamma_m) \right \rbrack + \frac{1}{5}\left \lbrack g(\gamma_n)-g(0)\right \rbrack, n= 1,2,3,4.
\end{equation}
In this paper, we study bifurcation mechanisms of the emergence of chaotic attractors and their types in this four-dimensional system.


\begin{figure*}[h]
\center{\includegraphics[width=0.99\linewidth]{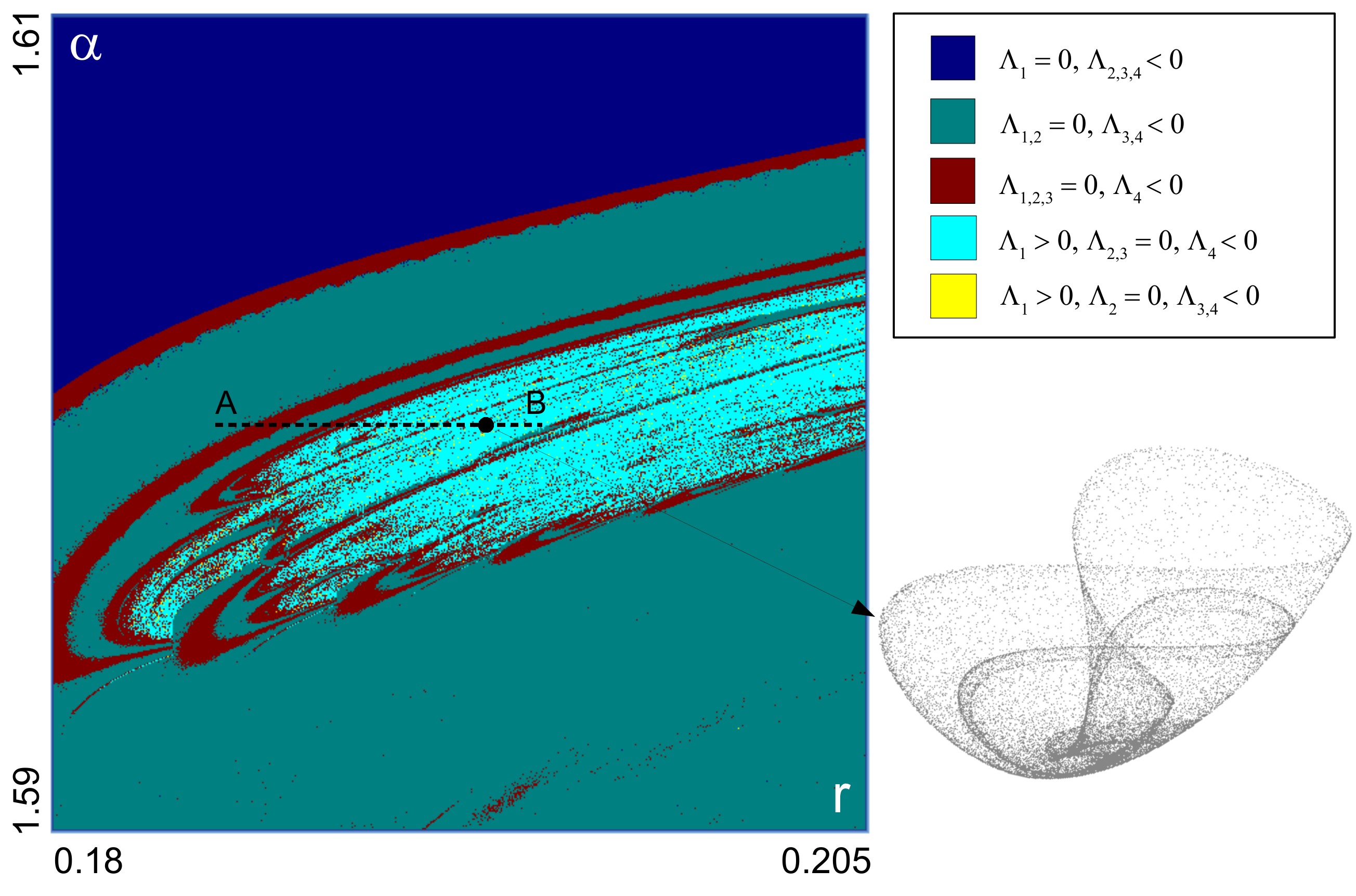} }
\vspace{-0.7cm}
\caption{{\footnotesize Lyapunov diagram for system \eqref{eq_mainEq}.}}
\label{fig1}
\end{figure*}

We start our studies with calculation of the Lyapunov diagram on the $(r,\alpha)$-parameter plane for $\beta=-1.58$, see Figure \ref{fig1}. For calculating Lyapunov exponents $\Lambda_1 \geq \Lambda_2 \geq \Lambda_3 > \Lambda_4$ we use the standard scheme \cite{BGGS80}. Depending on values of Lyapunov exponents each pixel of the resulting diagram is colored according to the palette presented in the right panel of Figure~\ref{fig1}. We took a simple stable regime -- a stable limit cycle -- and traced its evolution with changing parameters using also cross-section $\gamma_3 = 1.2$ (the same as in Ref.~\cite{AshTownOrosz2007}) and the corresponding 3D Poincar\'e map.

In accordance with the values of calculated Lyapunov exponents we have identified the following stable dynamical regimes:
\begin{itemize}
    \item $\Lambda_1 = 0, \Lambda_{2,3,4} < 0$ -- limit cycle (periodic point for the Poincar\'e map);
    \item $\Lambda_{1,2} = 0, \Lambda_{3,4} < 0$ -- quasiperiodic orbit (stable invariant curve for the Poincar\'e map) or nonhyperbolic periodic orbit;
    \item $\Lambda_{1,2,3} = 0, \Lambda_{4} < 0$ -- nonhyperbolic quasiperiodic orbit (nonhyperbolic invariant curve for the Poincar\'e map);
    \item $\Lambda_1 > 0, \Lambda_2 = 0, \Lambda_{3,4} < 0$ -- dissipative chaotic attractor;
    \item $\Lambda_1 > 0, \Lambda_{2,3} = 0, \Lambda_{4} < 0$ -- chaotic attractor with two zero Lyapunov exponents.
\end{itemize}
We note that in four-dimensional systems, hyperchaotic attractors with $\Lambda_1 > 0, \Lambda_2 > 0, \Lambda_3 = 0, \Lambda_{4} < 0$ and quasiperiodic regimes with $\Lambda_{1,2,3} = 0, \Lambda_{4} < 0$ are also possible, but in the system under consideration we do not observe these regimes. Here we consider the Lyapunov exponent to be indistinguishable from zero if it oscillates near zero within the range $5 \times 10^{-5}$.

Analysing the resulting Lyapunov diagram one can conclude that regions with chaotic attractors having the additional close to zero Lyapunov exponent prevail over other regions with chaotic behavior. An example of such attractors on the Poincar\'e map for $(r, \alpha) = (0.1933,1.60)$ is shown in the bottom right panel of Fig.~\ref{fig1}. It has the following set of Lyapunov exponents: 
$$
\Lambda_1 = 0.002179, \Lambda_2 = -0.000023, \Lambda_3 = -0.000008, \Lambda_4 = -0.028477,
$$
i.e., indeed a pair of its Lyapunov exponent $\Lambda_2$ and $\Lambda_3$ are near zero.

Lyapunov diagram presented in Fig.~\ref{fig1} also shows that strange attractors existing in system \eqref{eq_mainEq} cannot be robustly chaotic (pseudohyperbolic \cite{TS98,TS08,GKT21}) since regions with chaotic dynamics alternate with the so-called stability windows inside which simple (nonchaotic) dynamics are observed, i.e., in numerical experiments one can never be sure whether the chaotic attractor is observed or this is just a long transient process after which orbits tends to a simple regime. For most systems, a stable periodic orbit becomes an attractor inside a stability window. However, this is not the case of system \eqref{eq_mainEq}. The presence of stable quasiperiodic regimes inside stability is one more interesting and characteristic feature of this system. Moreover, we think that such organization of the parameter space is a typical feature for all systems demonstrating non-pseudohyperbolic attractors possessing the additional near-zero Lyapunov exponent.

Also let us note that the brown colored regions in Fig.~\ref{fig1} are actually lines on which bifurcations with quasiperiodic regime occur (period-doubling bifurcation or Neimark-Sacker bifurcation). However it is impossible to select the threshold of zero Lyapunov exponents in order to see this fact.

\section{Bifurcation of a fixed point with multipliers (1,1,1) and chaotic attractors with the additional zero Lyapunov exponent} \label{sec2}

\begin{figure*}[ht]
\center{\includegraphics[width=0.99\linewidth]{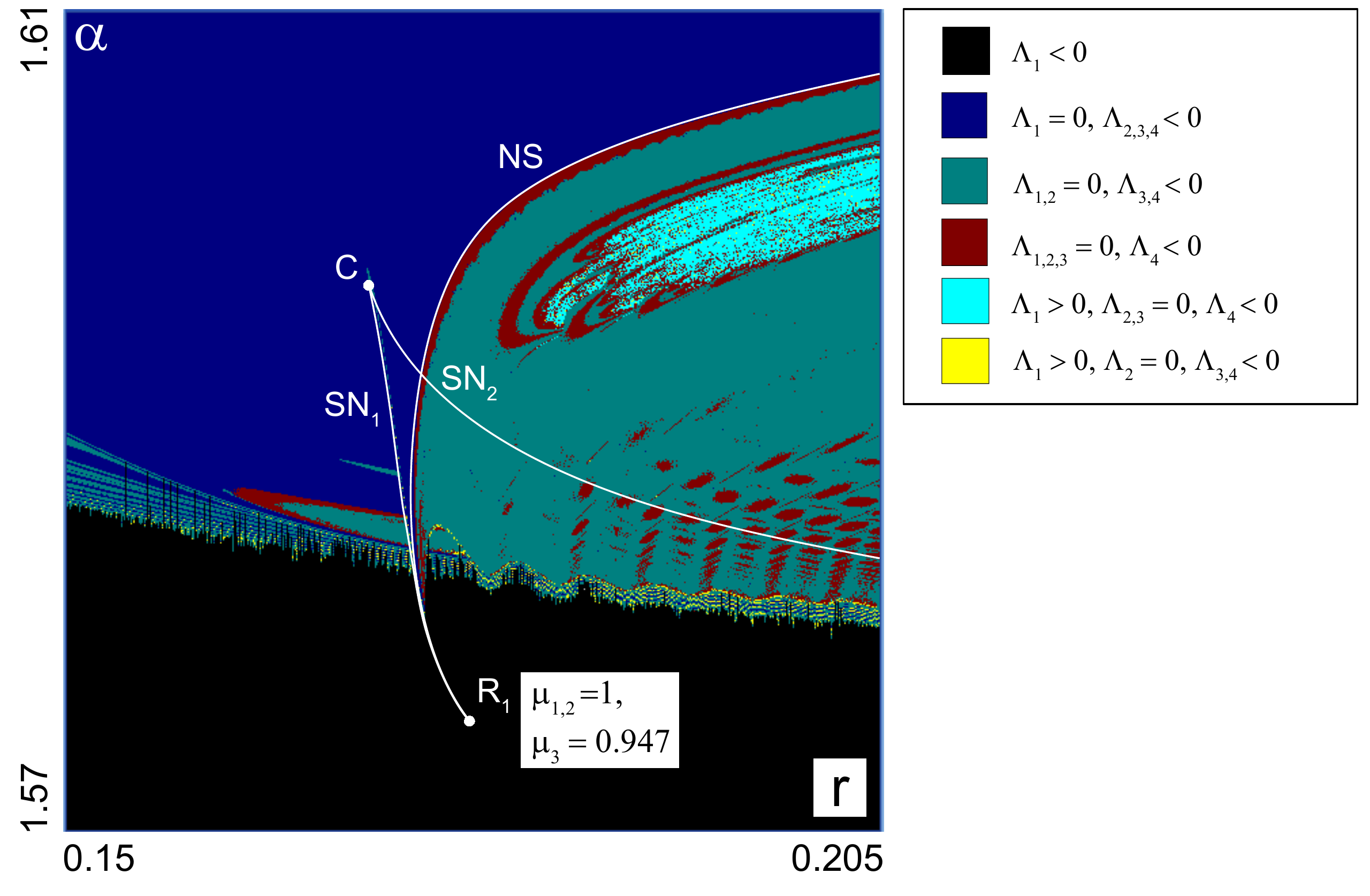} }
\vspace{-0.7cm}
\caption{{\footnotesize Lyapunov diagram superimposed with some bifurcation curves: SN$_1$ and SN$_2$ saddle-node bifurcations, NS -- supercritical Neimark-Sacker bifurcation, R$_1$ and C -- resonance 1:1 and cusp points.}}
\label{fig2}
\end{figure*}

In order to understand mechanisms of the formation of chaotic attractors with the additional near-zero Lyapunov exponent we perform bifurcation analysis on the same parameter plane as the Lyapunov diagram. The combined diagram for the extended parameter plane is shown in Figure~\ref{fig2}. All bifurcation curves are found with help of the MatCont package \cite{dhooge2008new, de2012interactive}.

A stable periodic orbit (fixed point) $O$ existing at the top-left part of the diagram undergoes a supercritical Neimark-Sacker bifurcation on the curve NS. As a result, this orbit becomes of a saddle-focus type and a stable invariant torus is born in its neighborhood. Another boundary of the stability region for fixed point $O$ is a saddle-node bifurcation curve SN$_1$. Crossing this curve from left to right one can observe the jump to another stable periodic orbit which is born below the second saddle-node bifurcation curve SN$_2$. The curve SN$_1$ merges with the SN$_2$-curve in the cusp point C and with the NS-curve in the $1:1$ resonance point R$_1$. 

\begin{figure}[h]
\center{\includegraphics[width=0.99\linewidth]{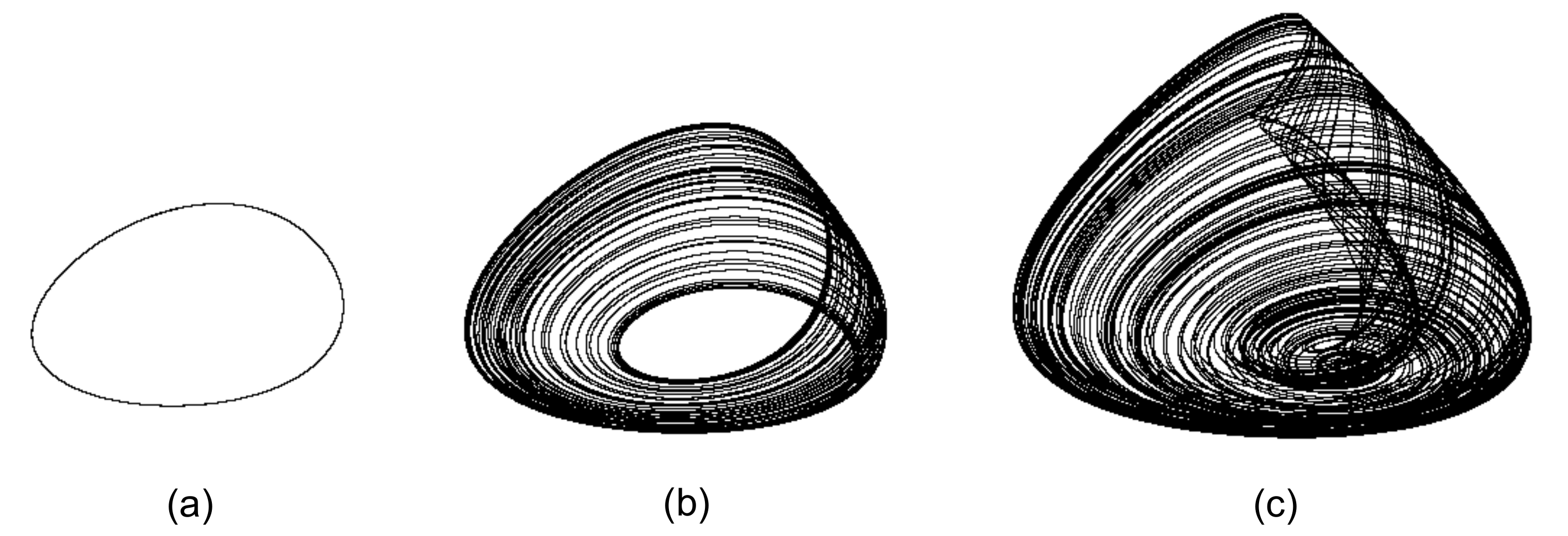} }
\vspace{-0.7cm}
\caption{{\footnotesize Main steps towards the appearance of Shilnikov attractor in system \eqref{eq_ACT} for $C=-0.4, B=-1$: (a) stable periodic orbit at $A=-0.7$, (b) R\"ossler-like attractor at $A=-0.8$, (c) Shilnikov spiral attractor at $A=-0.87$.}}
\label{fig3}
\end{figure}

At the point R$_1$ periodic orbit $O$ has a pair of unit multipliers $\mu_{1,2} = 1$. Its third multiplier is also close to one, $\mu_3 = 0.947$. Thus, the system near this point is close to the codimension-three bifurcation. The asymptotic normal form for this bifurcation, as well as for the bifurcation of an equilibrium state with three zero eigenvalues and Jordan block, coincides with the Arneod\'o-Coullet-Spiegel-Tresser system
\begin{equation}
\begin{cases}
\dot x = y, \\
\dot y = z, \\
\dot z = Ax + By + Cz + x^2.
\label{eq_ACT}
\end{cases}
\end{equation}
Bifurcations in this system were studied in detail in Refs.~\cite{ArnColTreSp1985,ArnColTreSp1985b}. In particular, it was shown that Shilnikov spiral attractors containing a saddle-focus equilibrium with a homoclinic loop exist in this system. In Refs.~\cite{GGKKB19,BKKKS20} it was explained how such attractors appear from the stable equilibrium via sequence of local and global bifurcations. At first, the stable equilibrium $P(0,0,0)$ undergoes a supercritical Andronov-Hopf bifurcation after which equilibrium $P$ becomes a saddle-focus and a stable periodic orbit is born in its neighborhood, see Fig.~\ref{fig3}a. Then, this periodic orbit goes through the cascade of period-doubling bifurcations resulting in a R\"ossler-like attractor, see Fig.~\ref{fig3}b. With further change of the governing parameter, orbits of the attractor pass closer and closer to the saddle-focus equilibrium and, finally, $P$ is absorbed by the attractor, due to the appearance of Shilnikov homoclinic loop, see Fig.~\ref{fig3}c. Lyapunov exponents for this attractor are:
$$
\Lambda_1 = 0.1017, \Lambda_2 = 0, \Lambda_3 = -0.5017,
$$
i.e., as any chaotic attractor of a system with continuous time, it has one zero Lyapunov exponent. Comparing phase portraits of the attractor found in system \eqref{eq_mainEq} (right panel in Fig.~\ref{fig1}) and the attractor presented in Fig.~\ref{fig3}c of one can note that these attractors are similar. An orbit taken in the first attractor looks like a discretization of a continuous orbit taken in the second attractor.

\begin{figure}[h]
\center{\includegraphics[width=0.99\linewidth]{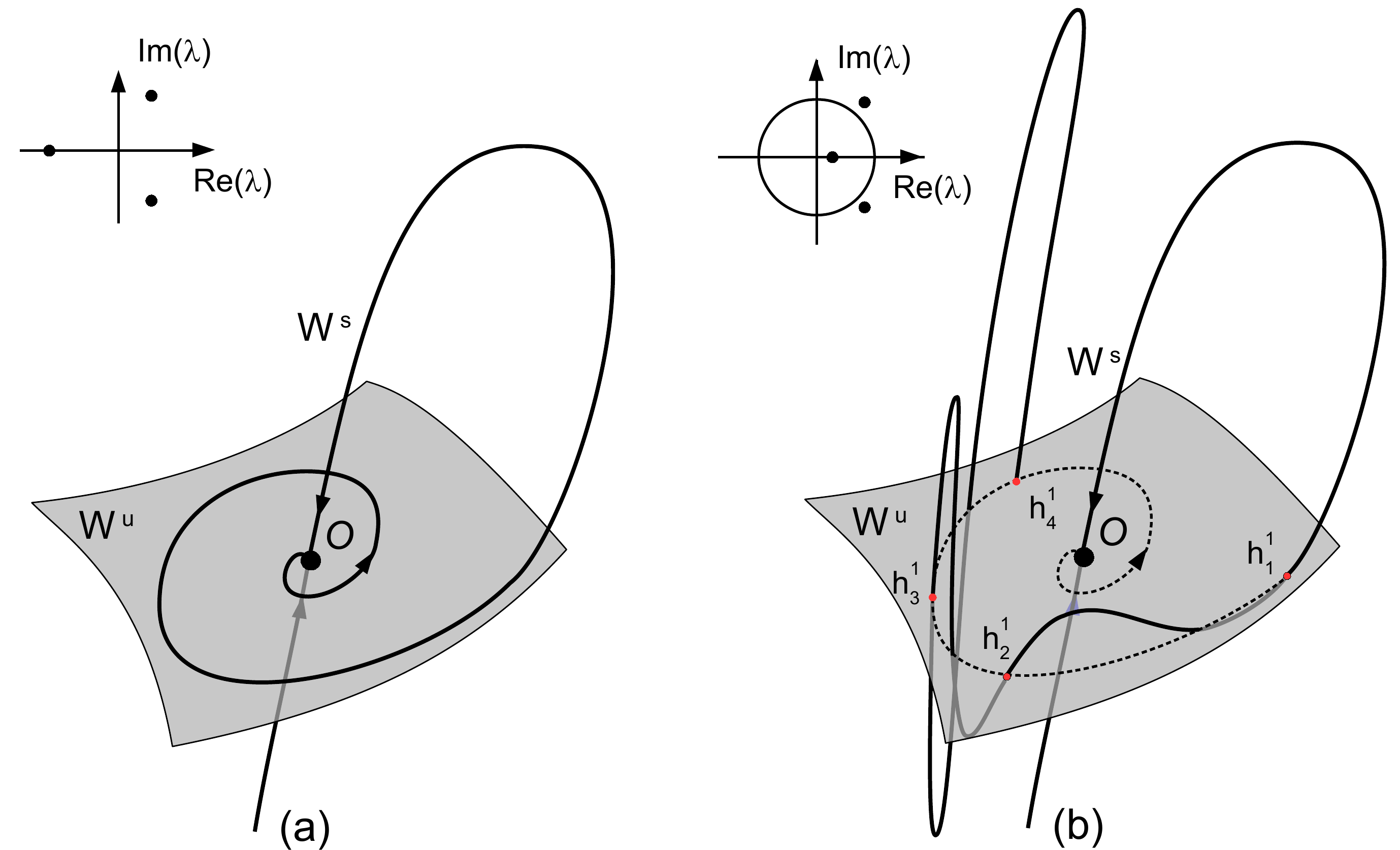} }
\vspace{-0.7cm}
\caption{{\footnotesize Homoclinic orbit to (a) saddle-focus equilibrium of some 3D ODE system, (b) saddle-focus fixed point for some 3D map. In both cases saddle-focus has the one-dimensional stable ($W^s$) and two-dimensional unstable ($W^u$) invariant manifolds. In the case of ODE system invariant manifold always intersect non-transversally, thus, the homoclinic orbit does not persist with changing in parameter values. In the case of map, the transversal interesection $W^s \cap W^u$ is shown. Points of this intersection - homoclinic points $h_1, h_2, \dots$ - persist for close maps.}}
\label{fig4}
\end{figure}

Chaotic attractors in system \eqref{eq_ACT}, as well as in many other systems (3D R\"ossler \cite{malykh2020homoclinic} system, Lotka–Volterra \cite{vano2006chaos} and Rosenzweig–MacArthur \cite{kuznetsov2001belyakov} models, etc.) appear in the full accordance with the Shilnikov scenario described in Ref~\cite{Shilnikov86}. The generalization of this scenario to the class of systems with discrete time (maps) was presented in Ref.~\cite{GonGonShil2012} (see also, Refs.~\cite{GonGonKazTur2014, GonGon2016}). In the framework of this scenario, bifurcations occur with a fixed point of the map (instead of the equilibrium state) and, further, with an invariant curve (instead of the limit cycle). In both cases the final bifurcation within the scenario is the formation of a homoclinic orbit to a saddle-focus. However, the homoclinic orbit to the equilibrium of any ODE system (see schematic example in Fig.~\ref{fig4}a) splits with arbitrarily small changes in parameter values: with inward splitting, new (double \cite{Gasp83}, triple \cite{GGNT97}, and so on) homoclinic orbits can appear, while with outward splitting new homoclinic orbits cannot appear since both unstable separatrices run away from an equilibrium. 

For maps another situation is observed in general case. A homoclinic orbit exists here in some open domains in the parameter space, since intersection between the stable ($W^s$) and unstable ($W^u$) invariant manifolds persists with changes in parameter values (of course, if these manifolds intersect transversely). Thus, discrete Shilnikov attractors, in principle, can exist in some open regions of the parameter space.

In the next section we study a scenario of the discrete Shilnikov attractor appearance in system \eqref{eq_mainEq} in one-parameter family and discuss a phenomenon when the invariant manifolds of the saddle-focus fixed point belonging to this attractor split under small changes in parameter values.

\section{Numerical evidences for the discrete Shilnikov attractor existence}
\label{sec3}

Here we study bifurcations leading from the stable periodic orbit $O$ (fixed point) to chaotic attractors in system \eqref{eq_mainEq} in the one-parameter family: $\alpha = 1.6,\, \beta = -1.58, \, r \in [0.175, 0.195]$ (pathway AB in Fig.~\ref{fig1}). All phase portraits are shown on the $(\gamma_1,\gamma_4)$-projections of 3D Poincar\'e map with the cross-section $\gamma_3=1.2$.

\begin{figure}[h]
\center{\includegraphics[width=0.99\linewidth]{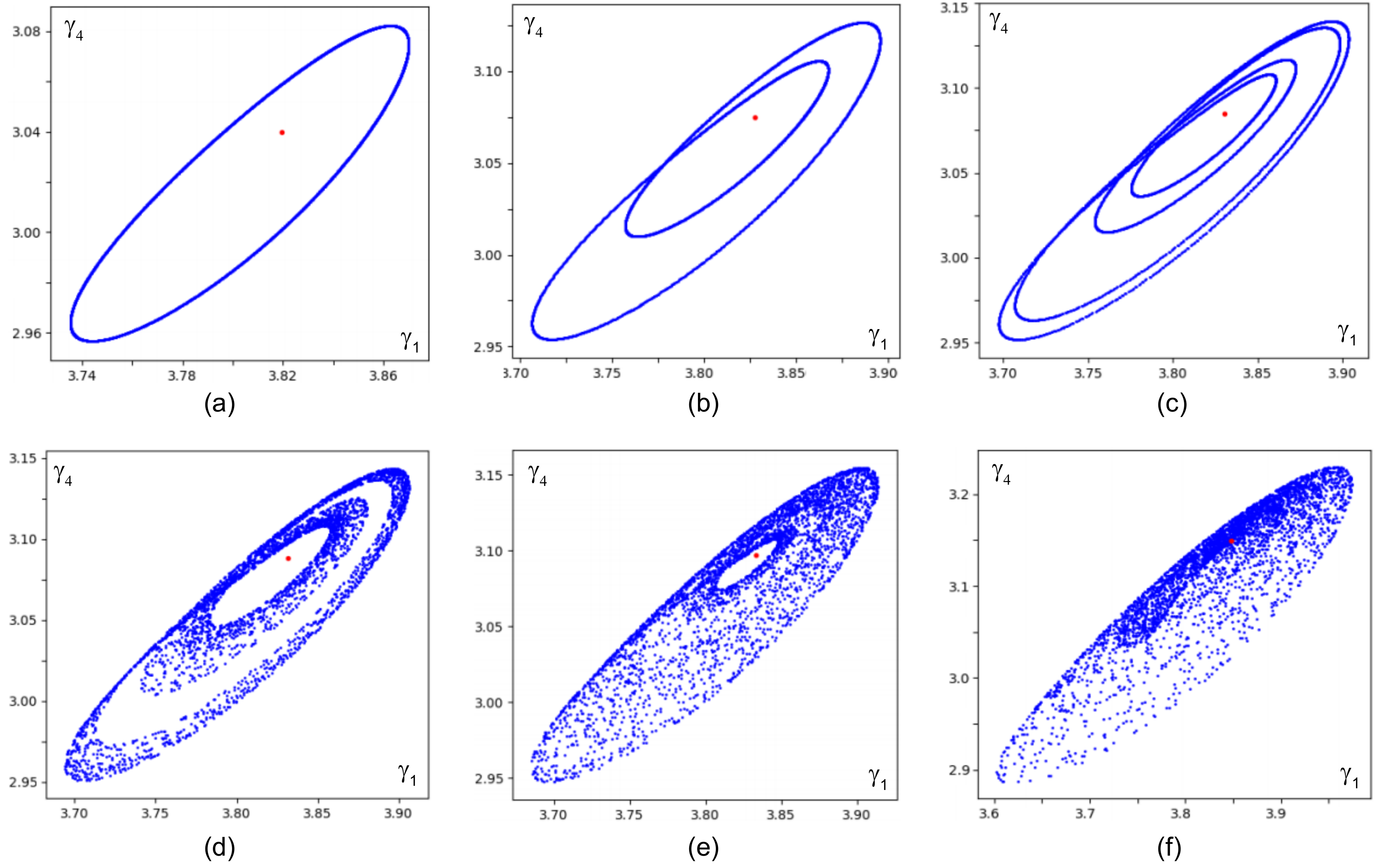} }
\vspace{-0.7cm}
\caption{{\footnotesize Projections of the Poincar\'e map of the attractor (blue-colored) and the saddle-focus fixed point $O$ (red-colored) on the plane $(\gamma_1, \gamma_4)$ along the pathway AB:  $\alpha = 1.6,\, \beta = -1.58, \, r \in [0.175, 0.195]$ (Fig.~\ref{fig1}): a) $r=0.185$, stable invariant curve; b) $r=0.188$, doubled invariant curve after the first period-doubling bifurcation; c) $r=0.1889$, four-round invariant curve after the two successful period-doubling bifurcations; d) $r=0.1892$, e) $r=0.19$, quasiperiodic H\'enon-like attractor \cite{broer2010chaos}; f) $r=0.195$, increasing $r$ leads to the decreasing in distance between the attractor and the fixed point $O$..}}
\label{fig5}
\end{figure}

For $r < 0.1795$, the stable fixed point $O$ is an attractor. At $r \approx 0.1795$ a supercritical Neimark-Sacker bifurcation occurs: the fixed point becomes a saddle-focus and a stable invariant curve $L$ appears in its neighborhood, Fig.~\ref{fig5}a. Right after this bifurcation the two-dimensional unstable manifold $W^u(O)$ is a disc with an edge on $L$. When we increase $r$ further, $W^u(O)$ starts to wind around the invariant curve $L$. After that we observe a sequence of period-doubling (length-doubling) bifurcation with invariant curve $L$, see more details about this bifurcations in Refs.~\cite{kaneko1983doubling,arneodo1983cascade,gonchenko2021doubling}. Figs.~\ref{fig5}b and \ref{fig5}c show the first and the second torus period-doubling bifurcations, respectively. 

\begin{figure}[!ht]
\begin{center}
\includegraphics[height=7.5cm]{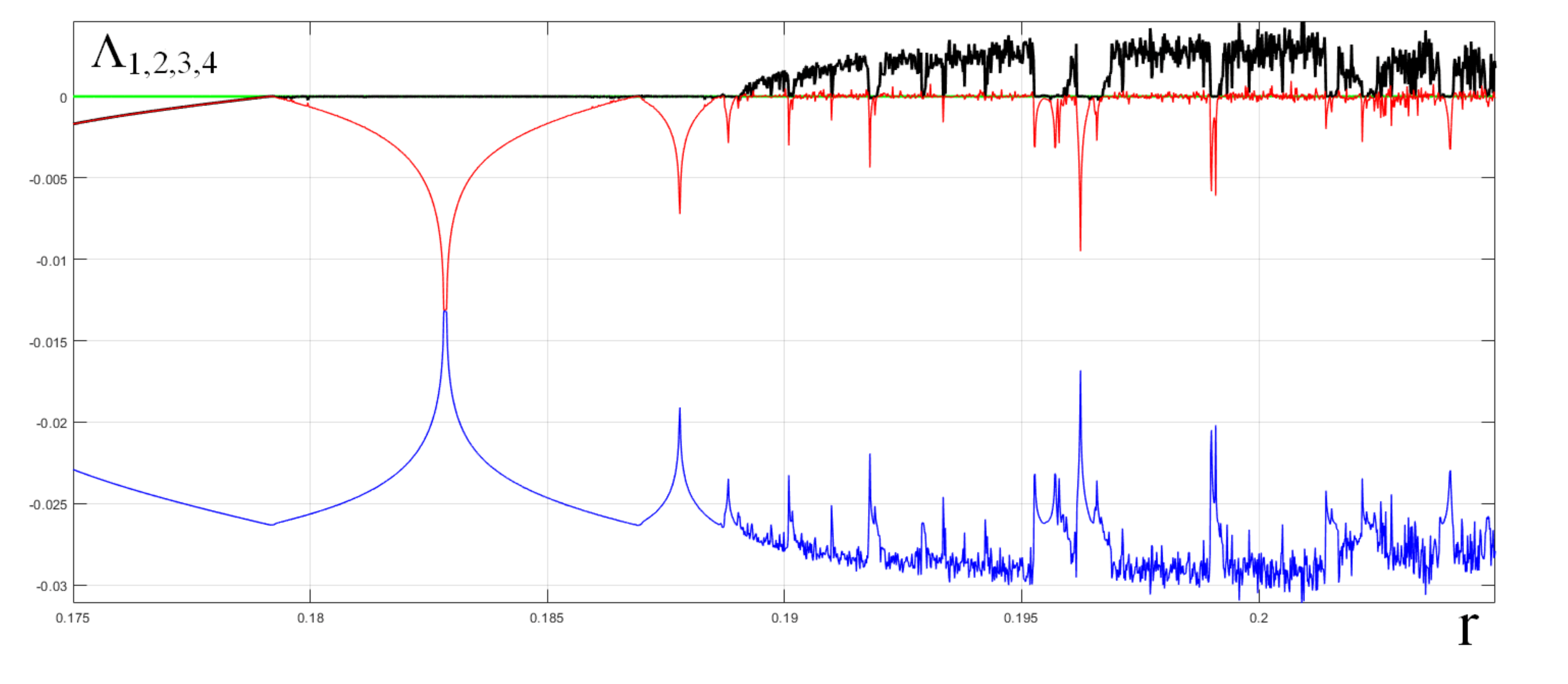}
\end{center}
\caption{Graph of Lyapunov exponents along the pathway AB}
\label{lyap-graph}
\end{figure}

After the sequence of period-doubling bifurcations we observe that the maximal Lyapunov exponent of the attractor becomes positive while next its two Lyapunov exponents are very close to zero, see the graph of Lyapunov exponents in Fig.~\ref{lyap-graph}. Phase portraits for some parameter values are shown in Figs.~\ref{fig5}d,e. In Refs. \cite{broer2002bifurcations, broer2005quasi, broer2010chaos} such attractors were called H\'enon-like quasiperiodic attractors by analogy with H\'enon-like attractors emerging after the cascades of period-doubling bifurcations with a fixed point and due to the fact that they have the additional near-zero Lyapunov exponent.

\begin{figure*}[h]
\center{\includegraphics[width=0.99\linewidth]{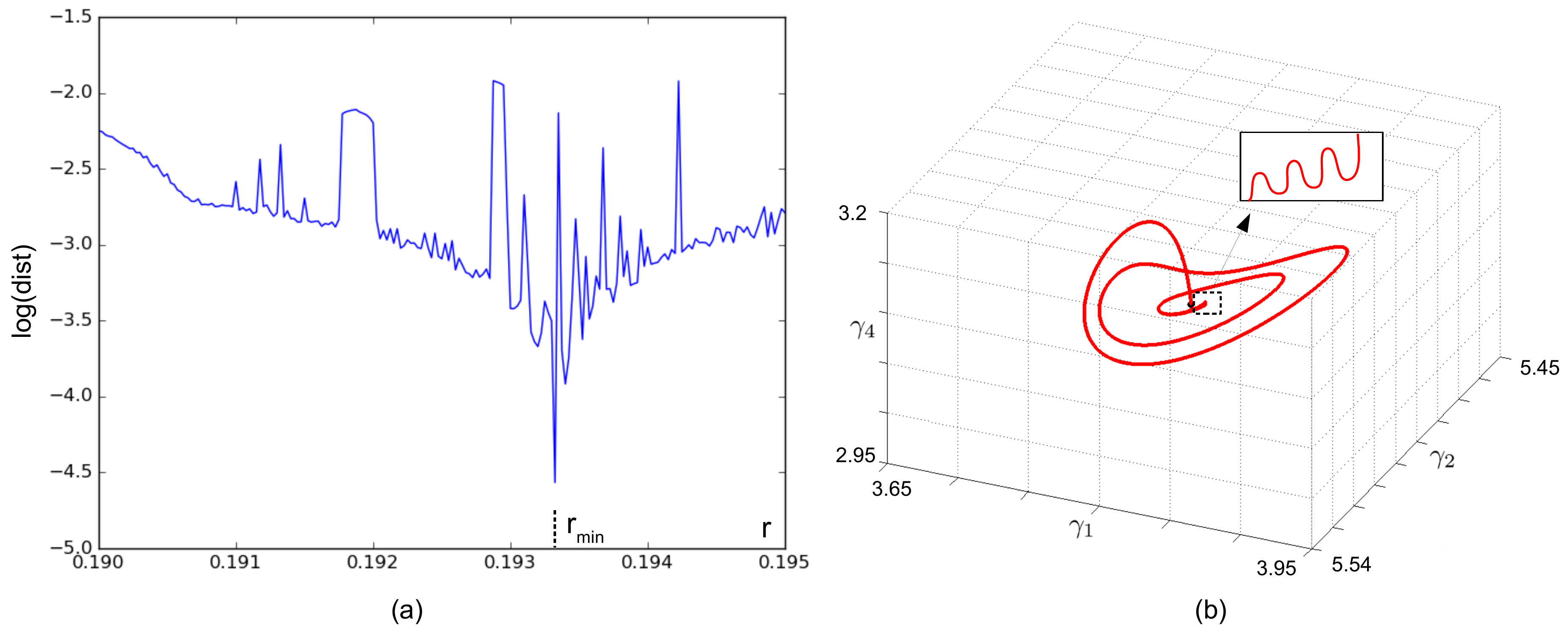} }
\vspace{-0.7cm}
\caption{{\footnotesize (a) Graph of distance between the saddle-focus fixed point $O$ and the attractor along the pathway AB; (b) a piece of the stable invariant manifold $W^s(O)$ (in the insert we show small  oscillations near $O$).}}
\label{fig7}
\end{figure*}

If we continue increasing parameter $r$ we observe that the chaotic attractor becomes closer and closer to the saddle-focus $O$ (visually, at least), Fig.~\ref{fig5}f. A graph of minimal distance between the attractor and $O$ on parameter $r$ presented in Fig. \ref{fig7}a confirms this claim and shows that at $r = r_{\rm min} \approx 0.1933365$ the distance almost vanishes. This feature indicates the appearance of a homoclinic intersection between $W^u(O)$ and $W^s(O)$, since here the fixed point $O$ with its two-dimensional unstable manifold $W^u(O)$ and homoclinic orbits seems to become a part of the attractor. We computed the stable invariant manifold $W^s(O)$ at $r = r_{\rm min}$ and found that in backward time it tends to $O$, see Fig.~\ref{fig7}b, i.e., we observe here the discrete Shilnikov attractor. It is interesting to note that oscillations of invariant manifold observing near the fixed point (typical for maps with homoclinic attractors) are almost invisible for the observed discrete Shilnikov attractor. Small changes in $r$ split the homoclinic orbit. As well as two indistinguishable from zero Lyapunov exponents on the observed chaotic attractors, we explain this phenomenon by the closeness of the 3D Poincar\'e map to the time-shift map of a 3D system of differential equations. 

\section{Discussion} \label{sec4}

In the present work we have studied chaotic dynamics observed in the 4D system of five identical globally coupled phase oscillators with the biharmonical coupling. We have shown that in this system there is a significant presence of chaotic attractors with additional zero Lyapunov exponent. Quite simple explanation for this phenomenon, based on the bifurcation analysis, is given. Interesting dynamics in this system is observed near the codimension three bifurcation, when the fixed point has the triplet of multipliers $(1,1,1)$. Therefore, the 3D Poincar\'e map for the system under consideration is close to a time shift map along orbits of the 3D system of differential equation which is an asymptotic normal form for this bifurcation.

Another interesting phenomenon discussed in this paper is a very weak splitting of invariant manifolds of the saddle-focus fixed point belonging to the observed discrete Shilnikov attractors. Such a phenomenon is also often met in systems of nonholonomic mechanics \cite{GGK13, BKS14, BorKazSat2016}. We explain it by the closeness of the Poincar\'e map for these systems to the time-shift map of systems of differential equations. Also we would like to underline one more interesting feature connected with the specific organization of bifurcation diagrams in regions with chaotic dynamics for systems with strange attractors possessing the additional near-zero Lyapunov exponent. Stable quasiperiodic regimes (ergodic invariant curves for the Poincar\'e map) populate stability windows in such system instead of usually observed periodic orbits for 3D or 4D systems with quasiattractors.

\begin{acknowledgments}
We thank S.V.~Gonchenko, G.V.~Osipov, and N.V.~Stankevich for fruitful discussions. This paper was supported by the Ministry of Science and Higher Education of Russian Federation (Project No. 0729-2020-0036). Numerical experiments in Sec.~\ref{sec2} were supported by the RSF grant No.~19-71-10048. The work of I.R.~Sataev was carried out within the framework of the state task of Kotel’nikov IRE RAS. A.~Kazakov also acknowledges financial support of the Theoretical Physics and Mathematics Advancement Foundation ``BASIS'' (Grant No. 20-7-1-36-5).
\end{acknowledgments}

\section*{Data Availability}
The data supporting numerical experiments presented in this paper are available from the corresponding author upon reasonable request.

\bibliography{refs}

\begin{thebibliography}{48}%
\makeatletter
\providecommand \@ifxundefined [1]{%
 \@ifx{#1\undefined}
}%
\providecommand \@ifnum [1]{%
 \ifnum #1\expandafter \@firstoftwo
 \else \expandafter \@secondoftwo
 \fi
}%
\providecommand \@ifx [1]{%
 \ifx #1\expandafter \@firstoftwo
 \else \expandafter \@secondoftwo
 \fi
}%
\providecommand \natexlab [1]{#1}%
\providecommand \enquote  [1]{``#1''}%
\providecommand \bibnamefont  [1]{#1}%
\providecommand \bibfnamefont [1]{#1}%
\providecommand \citenamefont [1]{#1}%
\providecommand \href@noop [0]{\@secondoftwo}%
\providecommand \href [0]{\begingroup \@sanitize@url \@href}%
\providecommand \@href[1]{\@@startlink{#1}\@@href}%
\providecommand \@@href[1]{\endgroup#1\@@endlink}%
\providecommand \@sanitize@url [0]{\catcode `\\12\catcode `\$12\catcode
  `\&12\catcode `\#12\catcode `\^12\catcode `\_12\catcode `\%12\relax}%
\providecommand \@@startlink[1]{}%
\providecommand \@@endlink[0]{}%
\providecommand \url  [0]{\begingroup\@sanitize@url \@url }%
\providecommand \@url [1]{\endgroup\@href {#1}{\urlprefix }}%
\providecommand \urlprefix  [0]{URL }%
\providecommand \Eprint [0]{\href }%
\providecommand \doibase [0]{http://dx.doi.org/}%
\providecommand \selectlanguage [0]{\@gobble}%
\providecommand \bibinfo  [0]{\@secondoftwo}%
\providecommand \bibfield  [0]{\@secondoftwo}%
\providecommand \translation [1]{[#1]}%
\providecommand \BibitemOpen [0]{}%
\providecommand \bibitemStop [0]{}%
\providecommand \bibitemNoStop [0]{.\EOS\space}%
\providecommand \EOS [0]{\spacefactor3000\relax}%
\providecommand \BibitemShut  [1]{\csname bibitem#1\endcsname}%
\let\auto@bib@innerbib\@empty
\bibitem [{\citenamefont {Gonchenko}, \citenamefont {Gonchenko},\ and\
  \citenamefont {Shilnikov}(2012)}]{GonGonShil2012}%
  \BibitemOpen
  \bibfield  {author} {\bibinfo {author} {\bibfnamefont {A.~S.}\ \bibnamefont
  {Gonchenko}}, \bibinfo {author} {\bibfnamefont {S.~V.}\ \bibnamefont
  {Gonchenko}}, \ and\ \bibinfo {author} {\bibfnamefont {L.~P.}\ \bibnamefont
  {Shilnikov}},\ }\bibfield  {title} {\enquote {\bibinfo {title} {Towards
  scenarios of chaos appearance in three-dimensional maps},}\ }\href@noop {}
  {\bibfield  {journal} {\bibinfo  {journal} {Rus. Nonlin. Dyn.}\ }\textbf
  {\bibinfo {volume} {8}},\ \bibinfo {pages} {3--28} (\bibinfo {year}
  {2012})}\BibitemShut {NoStop}%
\bibitem [{\citenamefont {Garashchuk}\ \emph {et~al.}(2019)\citenamefont
  {Garashchuk}, \citenamefont {Sinelshchikov}, \citenamefont {Kazakov},\ and\
  \citenamefont {Kudryashov}}]{GSKK19}%
  \BibitemOpen
  \bibfield  {author} {\bibinfo {author} {\bibfnamefont {I.~R.}\ \bibnamefont
  {Garashchuk}}, \bibinfo {author} {\bibfnamefont {D.~I.}\ \bibnamefont
  {Sinelshchikov}}, \bibinfo {author} {\bibfnamefont {A.~O.}\ \bibnamefont
  {Kazakov}}, \ and\ \bibinfo {author} {\bibfnamefont {N.~A.}\ \bibnamefont
  {Kudryashov}},\ }\bibfield  {title} {\enquote {\bibinfo {title} {Hyperchaos
  and multistability in the model of two interacting microbubble contrast
  agents},}\ }\href@noop {} {\bibfield  {journal} {\bibinfo  {journal} {Chaos:
  An Interdisciplinary Journal of Nonlinear Science}\ }\textbf {\bibinfo
  {volume} {29}},\ \bibinfo {pages} {063131} (\bibinfo {year}
  {2019})}\BibitemShut {NoStop}%
\bibitem [{\citenamefont {Pikovsky}\ and\ \citenamefont
  {Rosenblum}(2015)}]{PikRos2015}%
  \BibitemOpen
  \bibfield  {author} {\bibinfo {author} {\bibfnamefont {A.}~\bibnamefont
  {Pikovsky}}\ and\ \bibinfo {author} {\bibfnamefont {M.}~\bibnamefont
  {Rosenblum}},\ }\bibfield  {title} {\enquote {\bibinfo {title} {{Dynamics of
  globally coupled oscillators: Progress and perspectives}},}\ }\href {\doibase
  10.1063/1.4922971} {\bibfield  {journal} {\bibinfo  {journal} {Chaos: An
  Interdisciplinary Journal of Nonlinear Science}\ }\textbf {\bibinfo {volume}
  {25}},\ \bibinfo {pages} {097616} (\bibinfo {year} {2015})},\ \Eprint
  {http://arxiv.org/abs/https://doi.org/10.1063/1.4922971}
  {https://doi.org/10.1063/1.4922971} \BibitemShut {NoStop}%
\bibitem [{\citenamefont {Stankovski}\ \emph {et~al.}(2017)\citenamefont
  {Stankovski}, \citenamefont {Pereira}, \citenamefont {McClintock},\ and\
  \citenamefont {Stefanovska}}]{StankPerMcClintStefa2017}%
  \BibitemOpen
  \bibfield  {author} {\bibinfo {author} {\bibfnamefont {T.}~\bibnamefont
  {Stankovski}}, \bibinfo {author} {\bibfnamefont {T.}~\bibnamefont {Pereira}},
  \bibinfo {author} {\bibfnamefont {P.~V.~E.}\ \bibnamefont {McClintock}}, \
  and\ \bibinfo {author} {\bibfnamefont {A.}~\bibnamefont {Stefanovska}},\
  }\bibfield  {title} {\enquote {\bibinfo {title} {{Coupling functions:
  Universal insights into dynamical interaction mechanisms}},}\ }\href
  {\doibase 10.1103/RevModPhys.89.045001} {\bibfield  {journal} {\bibinfo
  {journal} {Rev. Mod. Phys.}\ }\textbf {\bibinfo {volume} {89}},\ \bibinfo
  {pages} {045001} (\bibinfo {year} {2017})}\BibitemShut {NoStop}%
\bibitem [{\citenamefont {Kuramoto}(1975)}]{Kur1975}%
  \BibitemOpen
  \bibfield  {author} {\bibinfo {author} {\bibfnamefont {Y.}~\bibnamefont
  {Kuramoto}},\ }\bibfield  {title} {\enquote {\bibinfo {title}
  {Self-entrainment of a population of coupled non-linear oscillators},}\ }in\
  \href@noop {} {\emph {\bibinfo {booktitle} {International symposium on
  mathematical problems in theoretical physics}}}\ (\bibinfo {organization}
  {Springer},\ \bibinfo {year} {1975})\ pp.\ \bibinfo {pages}
  {420--422}\BibitemShut {NoStop}%
\bibitem [{\citenamefont {Sakaguchi}\ and\ \citenamefont
  {Kuramoto}(1986)}]{KurSak1986}%
  \BibitemOpen
  \bibfield  {author} {\bibinfo {author} {\bibfnamefont {H.}~\bibnamefont
  {Sakaguchi}}\ and\ \bibinfo {author} {\bibfnamefont {Y.}~\bibnamefont
  {Kuramoto}},\ }\bibfield  {title} {\enquote {\bibinfo {title} {{A soluble
  active rotater model showing phase transitions via mutual entertainment}},}\
  }\href@noop {} {\bibfield  {journal} {\bibinfo  {journal} {Progress of
  Theoretical Physics}\ }\textbf {\bibinfo {volume} {76}},\ \bibinfo {pages}
  {576--581} (\bibinfo {year} {1986})}\BibitemShut {NoStop}%
\bibitem [{\citenamefont {Engelbrecht}\ and\ \citenamefont
  {Mirollo}(2014)}]{EngelMir2014}%
  \BibitemOpen
  \bibfield  {author} {\bibinfo {author} {\bibfnamefont {J.~R.}\ \bibnamefont
  {Engelbrecht}}\ and\ \bibinfo {author} {\bibfnamefont {R.}~\bibnamefont
  {Mirollo}},\ }\bibfield  {title} {\enquote {\bibinfo {title} {{Classification
  of attractors for systems of identical coupled Kuramoto oscillators}},}\
  }\href {\doibase 10.1063/1.4858458} {\bibfield  {journal} {\bibinfo
  {journal} {Chaos: An Interdisciplinary Journal of Nonlinear Science}\
  }\textbf {\bibinfo {volume} {24}},\ \bibinfo {pages} {013114} (\bibinfo
  {year} {2014})}\BibitemShut {NoStop}%
\bibitem [{\citenamefont {Okuda}(1993)}]{Okuda1993}%
  \BibitemOpen
  \bibfield  {author} {\bibinfo {author} {\bibfnamefont {K.}~\bibnamefont
  {Okuda}},\ }\bibfield  {title} {\enquote {\bibinfo {title} {Variety and
  generality of clustering in globally coupled oscillators},}\ }\href {\doibase
  https://doi.org/10.1016/0167-2789(93)90121-G} {\bibfield  {journal} {\bibinfo
   {journal} {Physica D: Nonlinear Phenomena}\ }\textbf {\bibinfo {volume}
  {63}},\ \bibinfo {pages} {424 -- 436} (\bibinfo {year} {1993})}\BibitemShut
  {NoStop}%
\bibitem [{\citenamefont {Hansel}, \citenamefont {Mato},\ and\ \citenamefont
  {Meunier}(1993)}]{HansMatoMeunier1993}%
  \BibitemOpen
  \bibfield  {author} {\bibinfo {author} {\bibfnamefont {D.}~\bibnamefont
  {Hansel}}, \bibinfo {author} {\bibfnamefont {G.}~\bibnamefont {Mato}}, \ and\
  \bibinfo {author} {\bibfnamefont {C.}~\bibnamefont {Meunier}},\ }\bibfield
  {title} {\enquote {\bibinfo {title} {Clustering and slow switching in
  globally coupled phase oscillators},}\ }\href {\doibase
  10.1103/PhysRevE.48.3470} {\bibfield  {journal} {\bibinfo  {journal} {Phys.
  Rev. E}\ }\textbf {\bibinfo {volume} {48}},\ \bibinfo {pages} {3470--3477}
  (\bibinfo {year} {1993})}\BibitemShut {NoStop}%
\bibitem [{\citenamefont {Kori}\ and\ \citenamefont
  {Kuramoto}(2001)}]{kori2001slow}%
  \BibitemOpen
  \bibfield  {author} {\bibinfo {author} {\bibfnamefont {H.}~\bibnamefont
  {Kori}}\ and\ \bibinfo {author} {\bibfnamefont {Y.}~\bibnamefont
  {Kuramoto}},\ }\bibfield  {title} {\enquote {\bibinfo {title} {Slow switching
  in globally coupled oscillators: robustness and occurrence through delayed
  coupling},}\ }\href@noop {} {\bibfield  {journal} {\bibinfo  {journal}
  {Physical Review E}\ }\textbf {\bibinfo {volume} {63}},\ \bibinfo {pages}
  {046214} (\bibinfo {year} {2001})}\BibitemShut {NoStop}%
\bibitem [{\citenamefont {Ashwin}\ \emph {et~al.}(2007)\citenamefont {Ashwin},
  \citenamefont {Orosz}, \citenamefont {Wordsworth},\ and\ \citenamefont
  {Townley}}]{AshTownOrosz2007}%
  \BibitemOpen
  \bibfield  {author} {\bibinfo {author} {\bibfnamefont {P.}~\bibnamefont
  {Ashwin}}, \bibinfo {author} {\bibfnamefont {G.}~\bibnamefont {Orosz}},
  \bibinfo {author} {\bibfnamefont {J.}~\bibnamefont {Wordsworth}}, \ and\
  \bibinfo {author} {\bibfnamefont {S.}~\bibnamefont {Townley}},\ }\bibfield
  {title} {\enquote {\bibinfo {title} {{Dynamics on Networks of Cluster States
  for Globally Coupled Phase Oscillators}},}\ }\href {\doibase
  10.1137/070683969} {\bibfield  {journal} {\bibinfo  {journal} {SIAM Journal
  on Applied Dynamical Systems}\ }\textbf {\bibinfo {volume} {6}},\ \bibinfo
  {pages} {728--758} (\bibinfo {year} {2007})},\ \Eprint
  {http://arxiv.org/abs/http://dx.doi.org/10.1137/070683969}
  {http://dx.doi.org/10.1137/070683969} \BibitemShut {NoStop}%
\bibitem [{\citenamefont {Popovych}, \citenamefont {Maistrenko},\ and\
  \citenamefont {Tass}(2005)}]{PopMaistrTass2005}%
  \BibitemOpen
  \bibfield  {author} {\bibinfo {author} {\bibfnamefont {O.~V.}\ \bibnamefont
  {Popovych}}, \bibinfo {author} {\bibfnamefont {Y.~L.}\ \bibnamefont
  {Maistrenko}}, \ and\ \bibinfo {author} {\bibfnamefont {P.~A.}\ \bibnamefont
  {Tass}},\ }\bibfield  {title} {\enquote {\bibinfo {title} {Phase chaos in
  coupled oscillators},}\ }\href {\doibase 10.1103/PhysRevE.71.065201}
  {\bibfield  {journal} {\bibinfo  {journal} {Phys. Rev. E}\ }\textbf {\bibinfo
  {volume} {71}},\ \bibinfo {pages} {065201} (\bibinfo {year}
  {2005})}\BibitemShut {NoStop}%
\bibitem [{\citenamefont {Bick}\ \emph {et~al.}(2011)\citenamefont {Bick},
  \citenamefont {Timme}, \citenamefont {Paulikat}, \citenamefont {Rathlev},\
  and\ \citenamefont {Ashwin}}]{AshBickTimm2011}%
  \BibitemOpen
  \bibfield  {author} {\bibinfo {author} {\bibfnamefont {C.}~\bibnamefont
  {Bick}}, \bibinfo {author} {\bibfnamefont {M.}~\bibnamefont {Timme}},
  \bibinfo {author} {\bibfnamefont {D.}~\bibnamefont {Paulikat}}, \bibinfo
  {author} {\bibfnamefont {D.}~\bibnamefont {Rathlev}}, \ and\ \bibinfo
  {author} {\bibfnamefont {P.}~\bibnamefont {Ashwin}},\ }\bibfield  {title}
  {\enquote {\bibinfo {title} {{Chaos in Symmetric Phase Oscillator
  Networks}},}\ }\href {\doibase 10.1103/PhysRevLett.107.244101} {\bibfield
  {journal} {\bibinfo  {journal} {Phys. Rev. Lett.}\ }\textbf {\bibinfo
  {volume} {107}},\ \bibinfo {pages} {244101} (\bibinfo {year}
  {2011})}\BibitemShut {NoStop}%
\bibitem [{\citenamefont {Bick}, \citenamefont {Ashwin},\ and\ \citenamefont
  {Rodrigues}(2016)}]{AshBickRodr2016}%
  \BibitemOpen
  \bibfield  {author} {\bibinfo {author} {\bibfnamefont {C.}~\bibnamefont
  {Bick}}, \bibinfo {author} {\bibfnamefont {P.}~\bibnamefont {Ashwin}}, \ and\
  \bibinfo {author} {\bibfnamefont {A.}~\bibnamefont {Rodrigues}},\ }\bibfield
  {title} {\enquote {\bibinfo {title} {Chaos in generically coupled phase
  oscillator networks with nonpairwise interactions},}\ }\href {\doibase
  10.1063/1.4958928} {\bibfield  {journal} {\bibinfo  {journal} {Chaos: An
  Interdisciplinary Journal of Nonlinear Science}\ }\textbf {\bibinfo {volume}
  {26}},\ \bibinfo {pages} {094814} (\bibinfo {year} {2016})},\ \Eprint
  {http://arxiv.org/abs/http://dx.doi.org/10.1063/1.4958928}
  {http://dx.doi.org/10.1063/1.4958928} \BibitemShut {NoStop}%
\bibitem [{\citenamefont {Broer}, \citenamefont {Sim{\'o}},\ and\ \citenamefont
  {Vitolo}(2002)}]{broer2002bifurcations}%
  \BibitemOpen
  \bibfield  {author} {\bibinfo {author} {\bibfnamefont {H.}~\bibnamefont
  {Broer}}, \bibinfo {author} {\bibfnamefont {C.}~\bibnamefont {Sim{\'o}}}, \
  and\ \bibinfo {author} {\bibfnamefont {R.}~\bibnamefont {Vitolo}},\
  }\bibfield  {title} {\enquote {\bibinfo {title} {{Bifurcations and strange
  attractors in the Lorenz-84 climate model with seasonal forcing}},}\
  }\href@noop {} {\bibfield  {journal} {\bibinfo  {journal} {Nonlinearity}\
  }\textbf {\bibinfo {volume} {15}},\ \bibinfo {pages} {1205} (\bibinfo {year}
  {2002})}\BibitemShut {NoStop}%
\bibitem [{\citenamefont {Broer}, \citenamefont {Vitolo},\ and\ \citenamefont
  {Sim{\'o}}(2005)}]{broer2005quasi}%
  \BibitemOpen
  \bibfield  {author} {\bibinfo {author} {\bibfnamefont {H.}~\bibnamefont
  {Broer}}, \bibinfo {author} {\bibfnamefont {R.}~\bibnamefont {Vitolo}}, \
  and\ \bibinfo {author} {\bibfnamefont {C.}~\bibnamefont {Sim{\'o}}},\
  }\bibfield  {title} {\enquote {\bibinfo {title} {{Quasi-periodic
  H{\'e}non-like attractors in the Lorenz-84 climate model with seasonal
  forcing}},}\ }in\ \href@noop {} {\emph {\bibinfo {booktitle} {EQUADIFF
  2003}}}\ (\bibinfo  {publisher} {World Scientific},\ \bibinfo {year} {2005})\
  pp.\ \bibinfo {pages} {601--606}\BibitemShut {NoStop}%
\bibitem [{\citenamefont {Gonchenko}\ \emph {et~al.}(2005)\citenamefont
  {Gonchenko}, \citenamefont {Ovsyannikov}, \citenamefont {Sim{\'o}},\ and\
  \citenamefont {Turaev}}]{GOST05}%
  \BibitemOpen
  \bibfield  {author} {\bibinfo {author} {\bibfnamefont {S.~V.}\ \bibnamefont
  {Gonchenko}}, \bibinfo {author} {\bibfnamefont {I.~I.}\ \bibnamefont
  {Ovsyannikov}}, \bibinfo {author} {\bibfnamefont {C.}~\bibnamefont
  {Sim{\'o}}}, \ and\ \bibinfo {author} {\bibfnamefont {D.~V.}\ \bibnamefont
  {Turaev}},\ }\bibfield  {title} {\enquote {\bibinfo {title}
  {{Three-dimensional H{\'e}non-like maps and wild Lorenz-like attractors}},}\
  }\href@noop {} {\bibfield  {journal} {\bibinfo  {journal} {International
  Journal of Bifurcation and Chaos}\ }\textbf {\bibinfo {volume} {15}},\
  \bibinfo {pages} {3493--3508} (\bibinfo {year} {2005})}\BibitemShut {NoStop}%
\bibitem [{\citenamefont {Broer}, \citenamefont {Sim{\'o}},\ and\ \citenamefont
  {Vitolo}(2010)}]{broer2010chaos}%
  \BibitemOpen
  \bibfield  {author} {\bibinfo {author} {\bibfnamefont {H.~W.}\ \bibnamefont
  {Broer}}, \bibinfo {author} {\bibfnamefont {C.}~\bibnamefont {Sim{\'o}}}, \
  and\ \bibinfo {author} {\bibfnamefont {R.}~\bibnamefont {Vitolo}},\
  }\bibfield  {title} {\enquote {\bibinfo {title} {Chaos and quasi-periodicity
  in diffeomorphisms of the solid torus},}\ }\href@noop {} {\bibfield
  {journal} {\bibinfo  {journal} {Discrete \& Continuous Dynamical Systems-B}\
  }\textbf {\bibinfo {volume} {14}},\ \bibinfo {pages} {871} (\bibinfo {year}
  {2010})}\BibitemShut {NoStop}%
\bibitem [{\citenamefont {Stankevich}\ \emph {et~al.}(2020)\citenamefont
  {Stankevich}, \citenamefont {Shchegoleva}, \citenamefont {Sataev},\ and\
  \citenamefont {Kuznetsov}}]{stankevich2020three}%
  \BibitemOpen
  \bibfield  {author} {\bibinfo {author} {\bibfnamefont {N.~V.}\ \bibnamefont
  {Stankevich}}, \bibinfo {author} {\bibfnamefont {N.~A.}\ \bibnamefont
  {Shchegoleva}}, \bibinfo {author} {\bibfnamefont {I.~R.}\ \bibnamefont
  {Sataev}}, \ and\ \bibinfo {author} {\bibfnamefont {A.~P.}\ \bibnamefont
  {Kuznetsov}},\ }\bibfield  {title} {\enquote {\bibinfo {title}
  {{Three-dimensional torus breakdown and chaos with two zero Lyapunov
  exponents in coupled radio-physical generators}},}\ }\href@noop {} {\bibfield
   {journal} {\bibinfo  {journal} {Journal of Computational and Nonlinear
  Dynamics}\ }\textbf {\bibinfo {volume} {15}} (\bibinfo {year}
  {2020})}\BibitemShut {NoStop}%
\bibitem [{\citenamefont {Shykhmamedov}\ \emph {et~al.}(2020)\citenamefont
  {Shykhmamedov}, \citenamefont {Karatetskaia}, \citenamefont {Kazakov},\ and\
  \citenamefont {Stankevich}}]{SKKS20}%
  \BibitemOpen
  \bibfield  {author} {\bibinfo {author} {\bibfnamefont {A.}~\bibnamefont
  {Shykhmamedov}}, \bibinfo {author} {\bibfnamefont {E.}~\bibnamefont
  {Karatetskaia}}, \bibinfo {author} {\bibfnamefont {A.}~\bibnamefont
  {Kazakov}}, \ and\ \bibinfo {author} {\bibfnamefont {N.}~\bibnamefont
  {Stankevich}},\ }\bibfield  {title} {\enquote {\bibinfo {title} {Hyperchaotic
  attractors of three-dimensional maps and scenarios of their appearance},}\
  }\href@noop {} {\bibfield  {journal} {\bibinfo  {journal} {arXiv preprint
  arXiv:2012.05099}\ } (\bibinfo {year} {2020})}\BibitemShut {NoStop}%
\bibitem [{\citenamefont {Karatetskaia}, \citenamefont {Shykhmamedov},\ and\
  \citenamefont {Kazakov}(2021)}]{KSK21}%
  \BibitemOpen
  \bibfield  {author} {\bibinfo {author} {\bibfnamefont {E.}~\bibnamefont
  {Karatetskaia}}, \bibinfo {author} {\bibfnamefont {A.}~\bibnamefont
  {Shykhmamedov}}, \ and\ \bibinfo {author} {\bibfnamefont {A.}~\bibnamefont
  {Kazakov}},\ }\bibfield  {title} {\enquote {\bibinfo {title} {Shilnikov
  attractors in three-dimensional orientation-reversing maps},}\ }\href@noop {}
  {\bibfield  {journal} {\bibinfo  {journal} {Chaos: An Interdisciplinary
  Journal of Nonlinear Science}\ }\textbf {\bibinfo {volume} {31}},\ \bibinfo
  {pages} {011102} (\bibinfo {year} {2021})}\BibitemShut {NoStop}%
\bibitem [{\citenamefont {Gonchenko}, \citenamefont {Gonchenko},\ and\
  \citenamefont {Kazakov}(2013)}]{GGK13}%
  \BibitemOpen
  \bibfield  {author} {\bibinfo {author} {\bibfnamefont {A.~S.}\ \bibnamefont
  {Gonchenko}}, \bibinfo {author} {\bibfnamefont {S.~V.}\ \bibnamefont
  {Gonchenko}}, \ and\ \bibinfo {author} {\bibfnamefont {A.~O.}\ \bibnamefont
  {Kazakov}},\ }\bibfield  {title} {\enquote {\bibinfo {title} {Richness of
  chaotic dynamics in nonholonomic models of a celtic stone},}\ }\href@noop {}
  {\bibfield  {journal} {\bibinfo  {journal} {Regular and Chaotic Dynamics}\
  }\textbf {\bibinfo {volume} {18}},\ \bibinfo {pages} {521--538} (\bibinfo
  {year} {2013})}\BibitemShut {NoStop}%
\bibitem [{\citenamefont {Borisov}, \citenamefont {Kazakov},\ and\
  \citenamefont {Sataev}(2016)}]{BorKazSat2016}%
  \BibitemOpen
  \bibfield  {author} {\bibinfo {author} {\bibfnamefont {A.~V.}\ \bibnamefont
  {Borisov}}, \bibinfo {author} {\bibfnamefont {A.~O.}\ \bibnamefont
  {Kazakov}}, \ and\ \bibinfo {author} {\bibfnamefont {I.~R.}\ \bibnamefont
  {Sataev}},\ }\bibfield  {title} {\enquote {\bibinfo {title} {{Spiral chaos in
  the nonholonomic model of a Chaplygin top}},}\ }\href {\doibase
  10.1134/S1560354716070157} {\bibfield  {journal} {\bibinfo  {journal}
  {Regular and Chaotic Dynamics}\ }\textbf {\bibinfo {volume} {21}},\ \bibinfo
  {pages} {939--954} (\bibinfo {year} {2016})}\BibitemShut {NoStop}%
\bibitem [{\citenamefont {Arneodo}\ \emph {et~al.}(1985)\citenamefont
  {Arneodo}, \citenamefont {Coullet}, \citenamefont {Spiegel},\ and\
  \citenamefont {Tresser}}]{ArnColTreSp1985}%
  \BibitemOpen
  \bibfield  {author} {\bibinfo {author} {\bibfnamefont {A.}~\bibnamefont
  {Arneodo}}, \bibinfo {author} {\bibfnamefont {P.}~\bibnamefont {Coullet}},
  \bibinfo {author} {\bibfnamefont {E.}~\bibnamefont {Spiegel}}, \ and\
  \bibinfo {author} {\bibfnamefont {C.}~\bibnamefont {Tresser}},\ }\bibfield
  {title} {\enquote {\bibinfo {title} {Asymptotic chaos},}\ }\href {\doibase
  https://doi.org/10.1016/0167-2789(85)90093-4} {\bibfield  {journal} {\bibinfo
   {journal} {Physica D: Nonlinear Phenomena}\ }\textbf {\bibinfo {volume}
  {14}},\ \bibinfo {pages} {327--347} (\bibinfo {year} {1985})}\BibitemShut
  {NoStop}%
\bibitem [{\citenamefont {Arneodo}, \citenamefont {Coullet},\ and\
  \citenamefont {Spiegel}(1985)}]{ArnColTreSp1985b}%
  \BibitemOpen
  \bibfield  {author} {\bibinfo {author} {\bibfnamefont {A.}~\bibnamefont
  {Arneodo}}, \bibinfo {author} {\bibfnamefont {P.}~\bibnamefont {Coullet}}, \
  and\ \bibinfo {author} {\bibfnamefont {E.}~\bibnamefont {Spiegel}},\
  }\bibfield  {title} {\enquote {\bibinfo {title} {The dynamics of triple
  convection},}\ }\href@noop {} {\bibfield  {journal} {\bibinfo  {journal}
  {Geophysical \& Astrophysical Fluid Dynamics}\ }\textbf {\bibinfo {volume}
  {31}},\ \bibinfo {pages} {1--48} (\bibinfo {year} {1985})}\BibitemShut
  {NoStop}%
\bibitem [{\citenamefont {Arneodo}, \citenamefont {Coullet},\ and\
  \citenamefont {Tresser}(1982)}]{ArnCoulTres1982}%
  \BibitemOpen
  \bibfield  {author} {\bibinfo {author} {\bibfnamefont {A.}~\bibnamefont
  {Arneodo}}, \bibinfo {author} {\bibfnamefont {P.}~\bibnamefont {Coullet}}, \
  and\ \bibinfo {author} {\bibfnamefont {C.}~\bibnamefont {Tresser}},\
  }\bibfield  {title} {\enquote {\bibinfo {title} {{Oscillators with chaotic
  behavior: An illustration of a theorem by Shil'nikov}},}\ }\href {\doibase
  10.1007/bf01011745} {\bibfield  {journal} {\bibinfo  {journal} {Journal of
  Statistical Physics}\ }\textbf {\bibinfo {volume} {27}},\ \bibinfo {pages}
  {171--182} (\bibinfo {year} {1982})}\BibitemShut {NoStop}%
\bibitem [{\citenamefont {Gonchenko}\ \emph {et~al.}(2014)\citenamefont
  {Gonchenko}, \citenamefont {Gonchenko}, \citenamefont {Kazakov},\ and\
  \citenamefont {Turaev}}]{GonGonKazTur2014}%
  \BibitemOpen
  \bibfield  {author} {\bibinfo {author} {\bibfnamefont {A.}~\bibnamefont
  {Gonchenko}}, \bibinfo {author} {\bibfnamefont {S.}~\bibnamefont
  {Gonchenko}}, \bibinfo {author} {\bibfnamefont {A.}~\bibnamefont {Kazakov}},
  \ and\ \bibinfo {author} {\bibfnamefont {D.}~\bibnamefont {Turaev}},\
  }\bibfield  {title} {\enquote {\bibinfo {title} {Simple scenarios of onset of
  chaos in three-dimensional maps},}\ }\href@noop {} {\bibfield  {journal}
  {\bibinfo  {journal} {International Journal of Bifurcation and Chaos}\
  }\textbf {\bibinfo {volume} {24}},\ \bibinfo {pages} {1440005} (\bibinfo
  {year} {2014})}\BibitemShut {NoStop}%
\bibitem [{\citenamefont {Gonchenko}\ and\ \citenamefont
  {Gonchenko}(2016)}]{GonGon2016}%
  \BibitemOpen
  \bibfield  {author} {\bibinfo {author} {\bibfnamefont {A.~S.}\ \bibnamefont
  {Gonchenko}}\ and\ \bibinfo {author} {\bibfnamefont {S.~V.}\ \bibnamefont
  {Gonchenko}},\ }\bibfield  {title} {\enquote {\bibinfo {title} {{Variety of
  strange pseudohyperbolic attractors in three-dimensional generalized
  H{\'e}non maps}},}\ }\href@noop {} {\bibfield  {journal} {\bibinfo  {journal}
  {Physica D: Nonlinear Phenomena}\ }\textbf {\bibinfo {volume} {337}},\
  \bibinfo {pages} {43--57} (\bibinfo {year} {2016})}\BibitemShut {NoStop}%
\bibitem [{\citenamefont {Grines}, \citenamefont {Kazakov},\ and\ \citenamefont
  {Sataev}(2017)}]{GrinKazSat2017}%
  \BibitemOpen
  \bibfield  {author} {\bibinfo {author} {\bibfnamefont {E.~A.}\ \bibnamefont
  {Grines}}, \bibinfo {author} {\bibfnamefont {A.~O.}\ \bibnamefont {Kazakov}},
  \ and\ \bibinfo {author} {\bibfnamefont {I.~R.}\ \bibnamefont {Sataev}},\
  }\bibfield  {title} {\enquote {\bibinfo {title} {{Discrete Shilnikov
  attractor and chaotic dynamics in the system of five identical globally
  coupled phase oscillators with biharmonic coupling}},}\ }\href {\doibase
  10.48550/ARXIV.1712.03839} {\bibfield  {journal} {\bibinfo  {journal}
  {arXiv}\ } (\bibinfo {year} {2017}),\ 10.48550/ARXIV.1712.03839},\ \Eprint
  {http://arxiv.org/abs/1712.03839} {arXiv:1712.03839} \BibitemShut {NoStop}%
\bibitem [{\citenamefont {Ashwin}\ and\ \citenamefont
  {Swift}(1992)}]{AshSwift1992}%
  \BibitemOpen
  \bibfield  {author} {\bibinfo {author} {\bibfnamefont {P.}~\bibnamefont
  {Ashwin}}\ and\ \bibinfo {author} {\bibfnamefont {J.~W.}\ \bibnamefont
  {Swift}},\ }\bibfield  {title} {\enquote {\bibinfo {title} {{The dynamics of
  $n$ weakly coupled identical oscillators}},}\ }\href {\doibase
  10.1007/BF02429852} {\bibfield  {journal} {\bibinfo  {journal} {Journal of
  Nonlinear Science}\ }\textbf {\bibinfo {volume} {2}},\ \bibinfo {pages}
  {69--108} (\bibinfo {year} {1992})}\BibitemShut {NoStop}%
\bibitem [{\citenamefont {Benettin}\ \emph {et~al.}(1980)\citenamefont
  {Benettin}, \citenamefont {Galgani}, \citenamefont {Giorgilli},\ and\
  \citenamefont {Strelcyn}}]{BGGS80}%
  \BibitemOpen
  \bibfield  {author} {\bibinfo {author} {\bibfnamefont {G.}~\bibnamefont
  {Benettin}}, \bibinfo {author} {\bibfnamefont {L.}~\bibnamefont {Galgani}},
  \bibinfo {author} {\bibfnamefont {A.}~\bibnamefont {Giorgilli}}, \ and\
  \bibinfo {author} {\bibfnamefont {J.-M.}\ \bibnamefont {Strelcyn}},\
  }\bibfield  {title} {\enquote {\bibinfo {title} {Lyapunov characteristic
  exponents for smooth dynamical systems and for {H}amiltonian systems; a
  method for computing all of them. part 1: Theory},}\ }\href@noop {}
  {\bibfield  {journal} {\bibinfo  {journal} {Meccanica}\ }\textbf {\bibinfo
  {volume} {15}},\ \bibinfo {pages} {9--20} (\bibinfo {year}
  {1980})}\BibitemShut {NoStop}%
\bibitem [{\citenamefont {Turaev}\ and\ \citenamefont
  {Shil'nikov}(1998)}]{TS98}%
  \BibitemOpen
  \bibfield  {author} {\bibinfo {author} {\bibfnamefont {D.~V.}\ \bibnamefont
  {Turaev}}\ and\ \bibinfo {author} {\bibfnamefont {L.~P.}\ \bibnamefont
  {Shil'nikov}},\ }\bibfield  {title} {\enquote {\bibinfo {title} {{An example
  of a wild strange attractor}},}\ }\href@noop {} {\bibfield  {journal}
  {\bibinfo  {journal} {Sbornik: Mathematics}\ }\textbf {\bibinfo {volume}
  {189}},\ \bibinfo {pages} {291--314} (\bibinfo {year} {1998})}\BibitemShut
  {NoStop}%
\bibitem [{\citenamefont {Turaev}\ and\ \citenamefont
  {Shil’nikov}(2008)}]{TS08}%
  \BibitemOpen
  \bibfield  {author} {\bibinfo {author} {\bibfnamefont {D.~V.}\ \bibnamefont
  {Turaev}}\ and\ \bibinfo {author} {\bibfnamefont {L.~P.}\ \bibnamefont
  {Shil’nikov}},\ }\bibfield  {title} {\enquote {\bibinfo {title}
  {{Pseudohyperbolicity and the problem on periodic perturbations of
  Lorenz-type attractors}},}\ }\href@noop {} {\bibfield  {journal} {\bibinfo
  {journal} {{Doklady Mathematics}}\ }\textbf {\bibinfo {volume} {77}},\
  \bibinfo {pages} {17--21} (\bibinfo {year} {2008})}\BibitemShut {NoStop}%
\bibitem [{\citenamefont {Gonchenko}, \citenamefont {Kazakov},\ and\
  \citenamefont {Turaev}(2021)}]{GKT21}%
  \BibitemOpen
  \bibfield  {author} {\bibinfo {author} {\bibfnamefont {S.}~\bibnamefont
  {Gonchenko}}, \bibinfo {author} {\bibfnamefont {A.}~\bibnamefont {Kazakov}},
  \ and\ \bibinfo {author} {\bibfnamefont {D.}~\bibnamefont {Turaev}},\
  }\bibfield  {title} {\enquote {\bibinfo {title} {Wild pseudohyperbolic
  attractor in a four-dimensional lorenz system},}\ }\href@noop {} {\bibfield
  {journal} {\bibinfo  {journal} {Nonlinearity}\ }\textbf {\bibinfo {volume}
  {34}},\ \bibinfo {pages} {2018} (\bibinfo {year} {2021})}\BibitemShut
  {NoStop}%
\bibitem [{\citenamefont {Dhooge}\ \emph {et~al.}(2008)\citenamefont {Dhooge},
  \citenamefont {Govaerts}, \citenamefont {Kuznetsov}, \citenamefont {Meijer},\
  and\ \citenamefont {Sautois}}]{dhooge2008new}%
  \BibitemOpen
  \bibfield  {author} {\bibinfo {author} {\bibfnamefont {A.}~\bibnamefont
  {Dhooge}}, \bibinfo {author} {\bibfnamefont {W.}~\bibnamefont {Govaerts}},
  \bibinfo {author} {\bibfnamefont {Y.~A.}\ \bibnamefont {Kuznetsov}}, \bibinfo
  {author} {\bibfnamefont {H.~G.~E.}\ \bibnamefont {Meijer}}, \ and\ \bibinfo
  {author} {\bibfnamefont {B.}~\bibnamefont {Sautois}},\ }\bibfield  {title}
  {\enquote {\bibinfo {title} {{New features of the software MatCont for
  bifurcation analysis of dynamical systems}},}\ }\href@noop {} {\bibfield
  {journal} {\bibinfo  {journal} {Mathematical and Computer Modelling of
  Dynamical Systems}\ }\textbf {\bibinfo {volume} {14}},\ \bibinfo {pages}
  {147--175} (\bibinfo {year} {2008})}\BibitemShut {NoStop}%
\bibitem [{\citenamefont {De~Witte}\ \emph {et~al.}(2012)\citenamefont
  {De~Witte}, \citenamefont {Govaerts}, \citenamefont {Kuznetsov},\ and\
  \citenamefont {Friedman}}]{de2012interactive}%
  \BibitemOpen
  \bibfield  {author} {\bibinfo {author} {\bibfnamefont {V.}~\bibnamefont
  {De~Witte}}, \bibinfo {author} {\bibfnamefont {W.}~\bibnamefont {Govaerts}},
  \bibinfo {author} {\bibfnamefont {Y.~A.}\ \bibnamefont {Kuznetsov}}, \ and\
  \bibinfo {author} {\bibfnamefont {M.}~\bibnamefont {Friedman}},\ }\bibfield
  {title} {\enquote {\bibinfo {title} {{Interactive initialization and
  continuation of homoclinic and heteroclinic orbits in MATLAB}},}\ }\href@noop
  {} {\bibfield  {journal} {\bibinfo  {journal} {ACM Transactions on
  Mathematical Software (TOMS)}\ }\textbf {\bibinfo {volume} {38}},\ \bibinfo
  {pages} {1--34} (\bibinfo {year} {2012})}\BibitemShut {NoStop}%
\bibitem [{\citenamefont {Gonchenko}\ \emph {et~al.}(2019)\citenamefont
  {Gonchenko}, \citenamefont {Gonchenko}, \citenamefont {Kazakov},
  \citenamefont {Kozlov},\ and\ \citenamefont {Bakhanova}}]{GGKKB19}%
  \BibitemOpen
  \bibfield  {author} {\bibinfo {author} {\bibfnamefont {S.~V.}\ \bibnamefont
  {Gonchenko}}, \bibinfo {author} {\bibfnamefont {A.~S.}\ \bibnamefont
  {Gonchenko}}, \bibinfo {author} {\bibfnamefont {A.~O.}\ \bibnamefont
  {Kazakov}}, \bibinfo {author} {\bibfnamefont {A.~D.}\ \bibnamefont {Kozlov}},
  \ and\ \bibinfo {author} {\bibfnamefont {Y.~V.}\ \bibnamefont {Bakhanova}},\
  }\bibfield  {title} {\enquote {\bibinfo {title} {{Mathematical theory of
  dynamical chaos and its applications: Review. Part 2. Spiral chaos of
  three-dimensional flows}},}\ }\href@noop {} {\bibfield  {journal} {\bibinfo
  {journal} {Izvestiya VUZ. Applied Nonlinear Dynamics}\ }\textbf {\bibinfo
  {volume} {27}},\ \bibinfo {pages} {7--52} (\bibinfo {year}
  {2019})}\BibitemShut {NoStop}%
\bibitem [{\citenamefont {Bakhanova}\ \emph {et~al.}(2020)\citenamefont
  {Bakhanova}, \citenamefont {Kazakov}, \citenamefont {Karatetskaya},
  \citenamefont {Kozlov},\ and\ \citenamefont {Saphonov}}]{BKKKS20}%
  \BibitemOpen
  \bibfield  {author} {\bibinfo {author} {\bibfnamefont {Y.~V.}\ \bibnamefont
  {Bakhanova}}, \bibinfo {author} {\bibfnamefont {A.~O.}\ \bibnamefont
  {Kazakov}}, \bibinfo {author} {\bibfnamefont {E.~Y.}\ \bibnamefont
  {Karatetskaya}}, \bibinfo {author} {\bibfnamefont {A.~D.}\ \bibnamefont
  {Kozlov}}, \ and\ \bibinfo {author} {\bibfnamefont {K.}~\bibnamefont
  {Saphonov}},\ }\bibfield  {title} {\enquote {\bibinfo {title} {On homoclinic
  attractors of three-dimensional flows},}\ }\href@noop {} {\bibfield
  {journal} {\bibinfo  {journal} {Izvestiya VUZ. Applied Nonlinear Dynamics}\
  }\textbf {\bibinfo {volume} {28}},\ \bibinfo {pages} {231--258} (\bibinfo
  {year} {2020})}\BibitemShut {NoStop}%
\bibitem [{\citenamefont {Malykh}\ \emph {et~al.}(2020)\citenamefont {Malykh},
  \citenamefont {Bakhanova}, \citenamefont {Kazakov}, \citenamefont
  {Pusuluri},\ and\ \citenamefont {Shilnikov}}]{malykh2020homoclinic}%
  \BibitemOpen
  \bibfield  {author} {\bibinfo {author} {\bibfnamefont {S.}~\bibnamefont
  {Malykh}}, \bibinfo {author} {\bibfnamefont {Y.}~\bibnamefont {Bakhanova}},
  \bibinfo {author} {\bibfnamefont {A.}~\bibnamefont {Kazakov}}, \bibinfo
  {author} {\bibfnamefont {K.}~\bibnamefont {Pusuluri}}, \ and\ \bibinfo
  {author} {\bibfnamefont {A.}~\bibnamefont {Shilnikov}},\ }\bibfield  {title}
  {\enquote {\bibinfo {title} {{Homoclinic chaos in the R{\"o}ssler model}},}\
  }\href@noop {} {\bibfield  {journal} {\bibinfo  {journal} {Chaos: An
  Interdisciplinary Journal of Nonlinear Science}\ }\textbf {\bibinfo {volume}
  {30}},\ \bibinfo {pages} {113126} (\bibinfo {year} {2020})}\BibitemShut
  {NoStop}%
\bibitem [{\citenamefont {Vano}\ \emph {et~al.}(2006)\citenamefont {Vano},
  \citenamefont {Wildenberg}, \citenamefont {Anderson}, \citenamefont {Noel},\
  and\ \citenamefont {Sprott}}]{vano2006chaos}%
  \BibitemOpen
  \bibfield  {author} {\bibinfo {author} {\bibfnamefont {J.}~\bibnamefont
  {Vano}}, \bibinfo {author} {\bibfnamefont {J.}~\bibnamefont {Wildenberg}},
  \bibinfo {author} {\bibfnamefont {M.}~\bibnamefont {Anderson}}, \bibinfo
  {author} {\bibfnamefont {J.}~\bibnamefont {Noel}}, \ and\ \bibinfo {author}
  {\bibfnamefont {J.}~\bibnamefont {Sprott}},\ }\bibfield  {title} {\enquote
  {\bibinfo {title} {{Chaos in low-dimensional Lotka--Volterra models of
  competition}},}\ }\href@noop {} {\bibfield  {journal} {\bibinfo  {journal}
  {Nonlinearity}\ }\textbf {\bibinfo {volume} {19}},\ \bibinfo {pages} {2391}
  (\bibinfo {year} {2006})}\BibitemShut {NoStop}%
\bibitem [{\citenamefont {Kuznetsov}, \citenamefont {De~Feo},\ and\
  \citenamefont {Rinaldi}(2001)}]{kuznetsov2001belyakov}%
  \BibitemOpen
  \bibfield  {author} {\bibinfo {author} {\bibfnamefont {Y.~A.}\ \bibnamefont
  {Kuznetsov}}, \bibinfo {author} {\bibfnamefont {O.}~\bibnamefont {De~Feo}}, \
  and\ \bibinfo {author} {\bibfnamefont {S.}~\bibnamefont {Rinaldi}},\
  }\bibfield  {title} {\enquote {\bibinfo {title} {{Belyakov homoclinic
  bifurcations in a tritrophic food chain model}},}\ }\href@noop {} {\bibfield
  {journal} {\bibinfo  {journal} {SIAM Journal on Applied Mathematics}\
  }\textbf {\bibinfo {volume} {62}},\ \bibinfo {pages} {462--487} (\bibinfo
  {year} {2001})}\BibitemShut {NoStop}%
\bibitem [{\citenamefont {Shilnikov}(1986)}]{Shilnikov86}%
  \BibitemOpen
  \bibfield  {author} {\bibinfo {author} {\bibfnamefont {L.~P.}\ \bibnamefont
  {Shilnikov}},\ }\bibfield  {title} {\enquote {\bibinfo {title} {Bifurcation
  theory and turbulence},}\ }\href@noop {} {\bibfield  {journal} {\bibinfo
  {journal} {Methods of the Qualitative Theory of Differential Equations,
  Gorky}\ ,\ \bibinfo {pages} {150--163}} (\bibinfo {year} {1986})}\BibitemShut
  {NoStop}%
\bibitem [{\citenamefont {Gaspard}(1983)}]{Gasp83}%
  \BibitemOpen
  \bibfield  {author} {\bibinfo {author} {\bibfnamefont {P.}~\bibnamefont
  {Gaspard}},\ }\bibfield  {title} {\enquote {\bibinfo {title} {Generation of a
  countable set of homoclinic flows through bifurcation},}\ }\href@noop {}
  {\bibfield  {journal} {\bibinfo  {journal} {Physics Letters A}\ }\textbf
  {\bibinfo {volume} {97}},\ \bibinfo {pages} {1--4} (\bibinfo {year}
  {1983})}\BibitemShut {NoStop}%
\bibitem [{\citenamefont {Gonchenko}\ \emph {et~al.}(1997)\citenamefont
  {Gonchenko}, \citenamefont {Turaev}, \citenamefont {Gaspard},\ and\
  \citenamefont {Nicolis}}]{GGNT97}%
  \BibitemOpen
  \bibfield  {author} {\bibinfo {author} {\bibfnamefont {S.~V.}\ \bibnamefont
  {Gonchenko}}, \bibinfo {author} {\bibfnamefont {D.~V.}\ \bibnamefont
  {Turaev}}, \bibinfo {author} {\bibfnamefont {P.}~\bibnamefont {Gaspard}}, \
  and\ \bibinfo {author} {\bibfnamefont {G.}~\bibnamefont {Nicolis}},\
  }\bibfield  {title} {\enquote {\bibinfo {title} {Complexity in the
  bifurcation structure of homoclinic loops to a saddle-focus},}\ }\href@noop
  {} {\bibfield  {journal} {\bibinfo  {journal} {Nonlinearity}\ }\textbf
  {\bibinfo {volume} {10}},\ \bibinfo {pages} {409} (\bibinfo {year}
  {1997})}\BibitemShut {NoStop}%
\bibitem [{\citenamefont {Kaneko}(1983)}]{kaneko1983doubling}%
  \BibitemOpen
  \bibfield  {author} {\bibinfo {author} {\bibfnamefont {K.}~\bibnamefont
  {Kaneko}},\ }\bibfield  {title} {\enquote {\bibinfo {title} {Doubling of
  torus},}\ }\href@noop {} {\bibfield  {journal} {\bibinfo  {journal} {Progress
  of theoretical physics}\ }\textbf {\bibinfo {volume} {69}},\ \bibinfo {pages}
  {1806--1810} (\bibinfo {year} {1983})}\BibitemShut {NoStop}%
\bibitem [{\citenamefont {Arneodo}, \citenamefont {Coullet},\ and\
  \citenamefont {Spiegel}(1983)}]{arneodo1983cascade}%
  \BibitemOpen
  \bibfield  {author} {\bibinfo {author} {\bibfnamefont {A.}~\bibnamefont
  {Arneodo}}, \bibinfo {author} {\bibfnamefont {P.}~\bibnamefont {Coullet}}, \
  and\ \bibinfo {author} {\bibfnamefont {E.~A.}\ \bibnamefont {Spiegel}},\
  }\bibfield  {title} {\enquote {\bibinfo {title} {Cascade of period doublings
  of tori},}\ }\href@noop {} {\bibfield  {journal} {\bibinfo  {journal}
  {Physics Letters A}\ }\textbf {\bibinfo {volume} {94}},\ \bibinfo {pages}
  {1--6} (\bibinfo {year} {1983})}\BibitemShut {NoStop}%
\bibitem [{\citenamefont {Gonchenko}, \citenamefont {Gonchenko},\ and\
  \citenamefont {Turaev}(2021)}]{gonchenko2021doubling}%
  \BibitemOpen
  \bibfield  {author} {\bibinfo {author} {\bibfnamefont {A.}~\bibnamefont
  {Gonchenko}}, \bibinfo {author} {\bibfnamefont {S.}~\bibnamefont
  {Gonchenko}}, \ and\ \bibinfo {author} {\bibfnamefont {D.}~\bibnamefont
  {Turaev}},\ }\bibfield  {title} {\enquote {\bibinfo {title} {Doubling of
  invariant curves and chaos in three-dimensional diffeomorphisms},}\
  }\href@noop {} {\bibfield  {journal} {\bibinfo  {journal} {Chaos: An
  Interdisciplinary Journal of Nonlinear Science}\ }\textbf {\bibinfo {volume}
  {31}},\ \bibinfo {pages} {113130} (\bibinfo {year} {2021})}\BibitemShut
  {NoStop}%
\bibitem [{\citenamefont {Borisov}, \citenamefont {Kazakov},\ and\
  \citenamefont {Sataev}(2014)}]{BKS14}%
  \BibitemOpen
  \bibfield  {author} {\bibinfo {author} {\bibfnamefont {A.~V.}\ \bibnamefont
  {Borisov}}, \bibinfo {author} {\bibfnamefont {A.~O.}\ \bibnamefont
  {Kazakov}}, \ and\ \bibinfo {author} {\bibfnamefont {I.~R.}\ \bibnamefont
  {Sataev}},\ }\bibfield  {title} {\enquote {\bibinfo {title} {{The reversal
  and chaotic attractor in the nonholonomic model of Chaplygin’s top}},}\
  }\href@noop {} {\bibfield  {journal} {\bibinfo  {journal} {Regular and
  Chaotic Dynamics}\ }\textbf {\bibinfo {volume} {19}},\ \bibinfo {pages}
  {718--733} (\bibinfo {year} {2014})}\BibitemShut {NoStop}%
\end{thebibliography}%

\end{document}